\documentclass[%
 reprint,
 superscriptaddress,
nofootinbib,
 amsmath,amssymb,
 aps,
 prd,
floatfix,
twocolumn,
notitlepage
]{revtex4-2}
\usepackage{graphicx}
\usepackage{dcolumn}
\usepackage{subfigure}
\usepackage{bm}
\usepackage{breakurl}
\usepackage{enumitem}
\usepackage{xcolor}
\usepackage[normalem]{ulem}

\begin{document}


\title{Complex networks with tuneable spectral dimension as a universality playground}

\author{Ana P. Mill\'an}
\affiliation{
Amsterdam UMC, Vrije Universiteit Amsterdam, Department of Clinical Neurophysiology and MEG Center, Amsterdam Neuroscience, De Boelelaan 1117, Amsterdam, The Netherlands}
\author{Giacomo Gori}
\affiliation{Institut f\"ur Theoretische Physik, Universit\"at 
Heidelberg, D-69120 Heidelberg, Germany}
\author{Federico Battiston}
\affiliation{Department of Network and Data Science, Central European University, 1051 Budapest, Hungary}
\author{Tilman Enss}
\affiliation{Institut f\"ur Theoretische Physik, Universit\"at 
Heidelberg, D-69120 Heidelberg, Germany}
\author{Nicol\`o Defenu}
\affiliation{Institute for Theoretical Physics, ETH Z\"urich
Wolfgang-Pauli-Str. 27, 8093 Zurich, Switzerland}
\affiliation{Institut f\"ur Theoretische Physik, Universit\"at 
Heidelberg, D-69120 Heidelberg, Germany}

\begin{abstract}
  Universality is one of the key concepts in understanding critical phenomena. 
  However, for interacting inhomogeneous systems described by complex networks a clear understanding of the relevant parameters for universality is still missing. 
  Here we discuss the role of a fundamental network parameter for universality, the spectral dimension. For this purpose, we construct a complex network model where the probability of a bond between two nodes is proportional to a power law of the nodes' distances. By explicit computation we prove that the spectral dimension for this model can be tuned continuously from $1$ to infinity, and we discuss related network connectivity measures. We propose our model as a tool to probe universal behaviour on inhomogeneous structures and comment on the possibility that the universal behaviour of correlated models on such networks mimics the one of continuous field theories in fractional Euclidean dimensions.
\end{abstract}

\maketitle

\section{Introduction} 
 
Scale invariance is a key property of critical systems and leads to the appearance of power-law scaling in several macroscopic physical quantities close to the transition point.  These power laws are universal as they appear in a large  variety of microscopically different systems\,\cite{Guggenheim1945}, which only share the presence of a symmetry breaking transition and the specific symmetry of the order parameter.  The existence of universality within the theory of critical phenomena was clarified several decades ago, thanks to the analogy between the thermodynamic limit of many body systems and the long-time behaviour of dynamical systems that was established by the renormalization group (RG) approach.  This success is exemplified by the study of phase transitions and spontaneous symmetry breaking in the paradigmatic $O(n)$-symmetric vector model\,\cite{Pellissetto2002}.  
The universal critical properties and scaling exponents depend only on the symmetry index $n$ and on the Euclidean spatial dimension $d$, which control the phase space for critical fluctuations.  With recent extensions to fractional $d,n\in\mathbb R$, this model describes universal critical phenomena in a wide range of physical systems\,\cite{ElShowk2014,Codello2013,Codello2015}.  
Going beyond the traditional case of thermal and quantum phase transitions\,\cite{Stanley1987,Sachdev:404196}, applications of universality include cell membranes\,\cite{Machta2012}, turbulence\,\cite{Alexakis2018}, fracture and plasticity\,\cite{Kardar1998, Shekhawat2013} and epidemics\,\cite{Cardy1985}.

Over the years, growing efforts have been devoted to mapping interacting systems and their complex patterns of connections into complex networks formed by a set of nodes and links describing their pairwise couplings~\cite{boccaletti2006complex}.  Indeed, networks are able to provide a useful abstraction to characterize the architecture of many real systems, on top of which collective behavior and criticality can emerge~\cite{dorogovtsev2008critical}.
 Comprehensive information on the structure of a network is provided by its spectrum~\cite{vanmieghembook}, in particular the one of the associated Laplacian~\cite{latorabook} for which many properties are known~\cite{delange2014Laplacian}.
 The graph Laplacian is known to be an important tool also to understand dynamics on networks, as it characterizes the return properties of the random walk~\cite{masuda2017random} and the stability of the synchronized state of a system of oscillators~\cite{boccaletti2006complex,arenas2008synchronization}. 
 
A quantity of particular interest is the \emph{spectral dimension} $d_s$, which characterizes the scaling of the eigenvalues of such Laplacian matrix\,\cite{rammal1984random,Burioni1996}. In fact, the traditional RG description of critical phenomena, where scaling behavior is influenced by diverging critical fluctuations, implicitly suggests the spectral dimension as the relevant control parameter for universal behavior on inhomogeneous structures. Long forgotten, this fundamental quantity has recently generated a new wave of interest\,\cite{bianconi2020spectral,torres2020simplicial,reitz2020higher} to characterize the structure of more complicated systems such as simplicial complexes, where couplings among constituents are not limited to pairwise interactions~\cite{battiston2020networks}.  For many complex networks, the Fiedler (second smallest) eigenvalue remains finite in the thermodynamic limit, in which case the network is said to display a \textit{spectral gap}. By contrast, if the spectral gap closes as the system size grows, the network is said to have a finite spectral dimension \cite{Burioni1996}.
 
The role of the spectral dimension as a control parameter for universal behavior in critical phenomena can be proven in quadratic models, such as the spherical model\,\cite{Joyce1966} and Dyson's hierarchical model\,\cite{Dyson1969,Meurice2007} in the mean field regime.  Its validity for correlated critical models has proven much harder to verify, despite several investigations on classical long-range systems\,\cite{Angelini2014, Defenu2015, Defenu:2017dc, Gori2017}, diluted models\,\cite{Leuzzi2013, Berganza2013, Cescatti2019}, spin glasses\,\cite{Banos2012, Katzgraber2009} and quantum systems\,\cite{Defenu:2017dc}.  Several of these investigations rely on a conjectured relation between the universality of long-range interacting systems and that of local models with $d\in\mathbb{R}$.
Such a relation could not be verified either by functional RG or by conformal bootstrap investigations, which instead confirmed its approximate nature\,\cite{Defenu2015, Behan2017, Behan2017b}.

In parallel, geometrical investigations of network structures have also considered the \textit{fractal dimension} $d_{f}$, which characterizes the scaling of the number of neighbors of a node $N_{n}$ as a function of distance\,\cite{albert2002statistical,gallos2012small}. Based on such a definition the fractal dimension depends on the metric employed, for instance
considering the network distance $\rho$ the fractal dimension $d_{f}$ is defined according to the relation $N_n(\rho) \sim \rho^{d_{f}}$. In the following we always imply such a definition, unless explicitly stated that we refer to the definition based on Euclidean distance. 
Traditional investigations of Ising models on fractals found a nontrivial dependence of the scaling exponents on the (Euclidean) fractal dimension $d_{f}$ when $1<d_{f}<2$\,\cite{Gefen1980,Gefen1984,Gefen1984b}. Subsequently, it was shown that the universal scaling on fractals was not uniquely determined by the fractal dimension, leading to claims of a universality breakdown for $d_{f}<2$\,\cite{Gefen1980}.  This claim, however, is misleading since there is the possibility that true universality is found as a function of the \emph{spectral} instead of the fractal dimension even for $d_f<2$, at least on a restricted class of graphs.

The intention of this work is to provide a suitable tool for future numerical investigations of the dependence of the universal properties on the spectral dimension. Therefore, we propose a non-weighted graph, whose nodes are arranged in such a way that the spectral dimension is finite and can be continuously tuned. The model is constructed in Sec.\,\ref{model} from a one-dimensional nearest-neighbour chain, where additional long-distance bonds are inserted randomly with a probability that decays as a power law of the bond length. The spectral dimension of the model is explicitly computed as a function of the power-law decay exponent $\sigma$. Besides, in order to prove the numerical stability of the model, it has been numerically verified that it controls both the scaling of the spectrum (Sec.\,\ref{spec}) and the return times of \emph{random walkers} (RWs) in Sec.\,\ref{RW} as predicted in Ref.\,\cite{Burioni1996}. 

In Sec.\,\ref{univ_pl}, we present our model as a platform to study universality and critical phenomena on complex networks. We characterise the return rate of random walks on this network by computing the anomalous dimension of the model, defined in analogy to previous studies of critical phenomena with long-range interactions\,\cite{Grassberger2013SIR, Grassberger20132DSIR,Gori2017}. Interestingly, a striking resemblance between this anomalous dimension and the one of the scalar $\varphi^{4}$ theory at non-integer values of the spectral dimension is found. Finally, in Sec.\,\ref{concl} we conclude with a discussion of the future perspectives of our findings.

\section{Model}
\label{model}
\begin{figure}[ht]
\centering
\includegraphics[width=0.85\columnwidth]{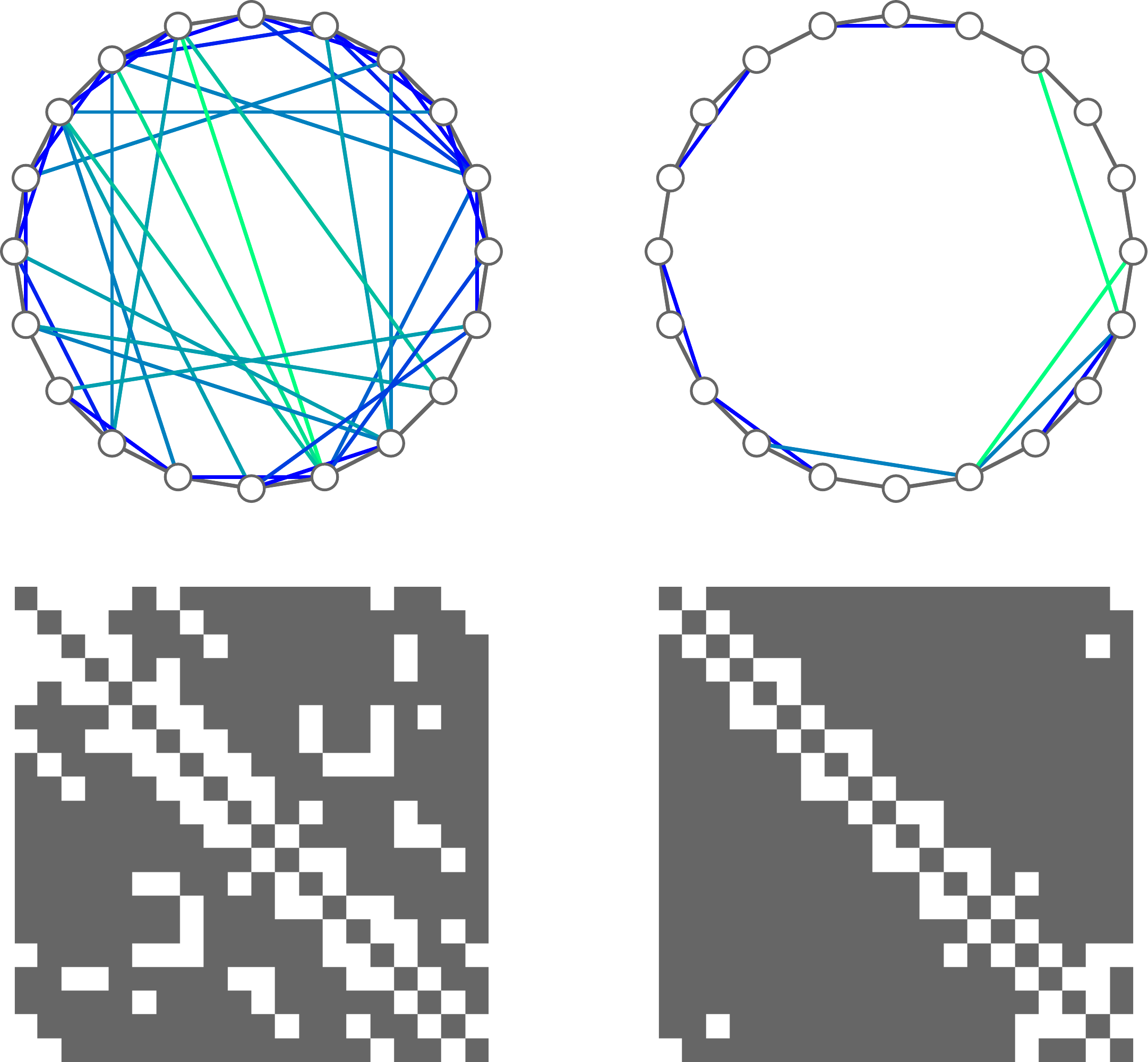}
\caption{Examples of the network layout for $N=20$, $\sigma=2/3,3/2$ (left) and (right), respectively. The adjacency matrices are shown in the bottom row, indicating existing edges (white squares). \label{Fig1} }
\end{figure}
We consider a network of $N$ nodes placed regularly on a circumference of radius $1$, at locations $\theta_i = 2\pi i/N$, $i=1,...,N$. 
The network is characterized by its adjacency matrix $A=\{a_{ij} \}$, where $a_{ij}\in\lbrace 0,1\rbrace$ indicates, respectively, the absence or presence of a link between nodes $i$ and $j$. The coupling probability between any pair of nodes is given by
\begin{equation}\label{eq:netw_def}
p_{ij} = \frac{1}{r_{ij}^{1+\sigma}},
\end{equation}
where $r_{ij}\neq 0$ is the distance between nodes $i$ and $j$ and $\sigma$ is the model parameter characterizing the  scaling of the coupling probability with the (geometric) distance. Note that our network does not contain self-loops, hence we only consider links with $i\neq j$.
Consequently, the model generates networks with tightly connected local neighbourhoods and increasingly rare long-range connections. As a robustness check for our results, we will consider two versions of the aforementioned model, based on different definitions of the distance 
\begin{align}
\label{l_def}
r^{(L)}_{ij}&=\min(|i-j|, N-|i-j|),\\
\label{c_def}
r^{(C)}_{ij}&=\left.\sin\left(\frac{\pi}{N}|i-j|\right)\right/\sin\left(\frac{\pi}{N}\right),
\end{align}
for the linear ($L$) and circular ($C$) models.

Long-range power-law decaying interactions have a long history in the study of critical phenomena and are known to influence the critical scaling behavior close to a second-order phase transition both at\,\cite{Dyson1969,Thouless1969,anderson1971some,fisher1972critical,aizenman1988critical,sak1973recursion} and out-of-equilibrium\,\cite{hinrichsen2007non}. More closely related to the present studies, extended couplings with power-law decaying probabilities played a crucial role in the investigations of epidemic processes, where they alter the  scaling of observables\,\cite{grassberger1986spreading,janessen1999levy, linder2008long,Grassberger2013SIR}. Along these lines, the network model proposed above can be regarded as the giant cluster of a long-range percolation model well inside the percolating regime\,\cite{Grassberger2013SIR, Grassberger20132DSIR,Gori2017}, such that this cluster always contains a nearest-neighbour connected ring backbone, $p_{i,i+1}=1\ \forall i$.  More recently,  the two-dimensional lattice version of this model has been employed to investigate critical dynamics in the XY model\,\cite{Expert2017,DeNigris2019} as well as epidemics spreading in $d=1$ and $2$\,\cite{Grassberger20132DSIR,Grassberger2013SIR}. Within the network theory context, the present model can be regarded as a one-dimensional instance of the Kleinberg model\,\cite{Kleinberg2000}, introduced to investigate the emergence of the small-world phenomenon beyond the paradigm of Watts and Strogatz\,\cite{watts1998collective} and  conventionally employed in the study of optimal transport problems\,\cite{Carmi2009,Li2010,Weng2015}. 
 
As mentioned above, the specific one dimensional model under study, which we refer to as long-range random ring (LRRR), was already employed to study the critical properties of long-range epidemics\,\cite{Grassberger2013SIR} and percolation\,\cite{Gori2017}. In the following, we are going to show that its spectral properties are highly non-trivial and realize the whole range of spectral dimensions $d_{s}\in [1,\infty)$. In particular, and in contrast to the two-dimensional version analyzed in Ref.\,\cite{Berganza2013}, our one-dimensional model allows to realize low spectral dimensions $d_{s}<2$, which are expected to be very relevant in the study of universal behavior for critical models with non-continuous symmetries, such as the Ising model and percolation\,\cite{Gefen1980, Gefen1984}. An example of our network layout and adjacency matrices for two values of $\sigma$ is shown in Fig.\,\ref{Fig1}. For simplicity, we will here consider undirected symmetric networks with $a_{ij}=a_{ji}$.

\begin{figure*}[ht!]
	\centering
	\subfigure[\hspace*{-2em}]{\label{Fig2a}\includegraphics[width=.32\textwidth]{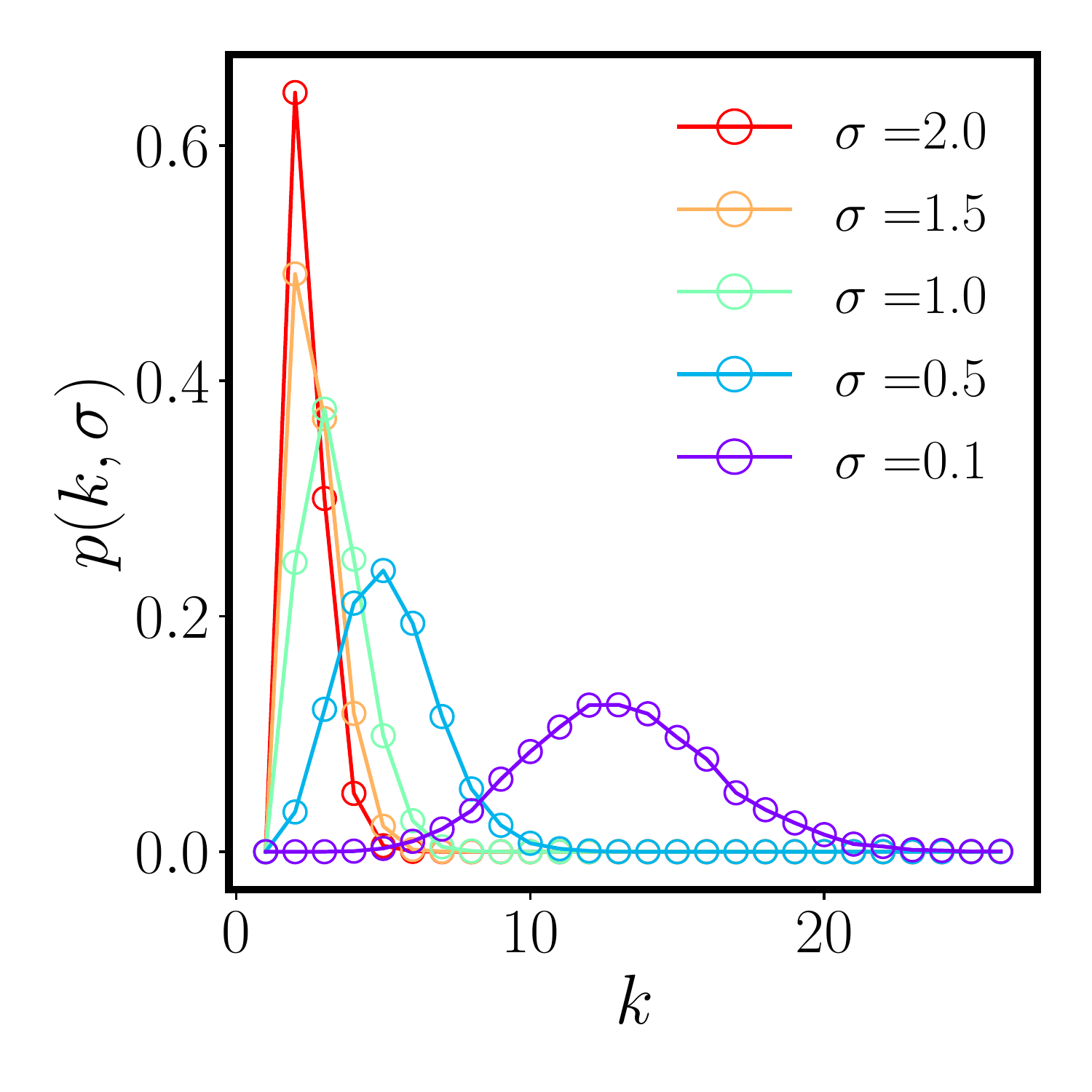}}
	\hfill
	\subfigure[\hspace*{-2em}]{\label{Fig2b}\includegraphics[width=.32\textwidth]{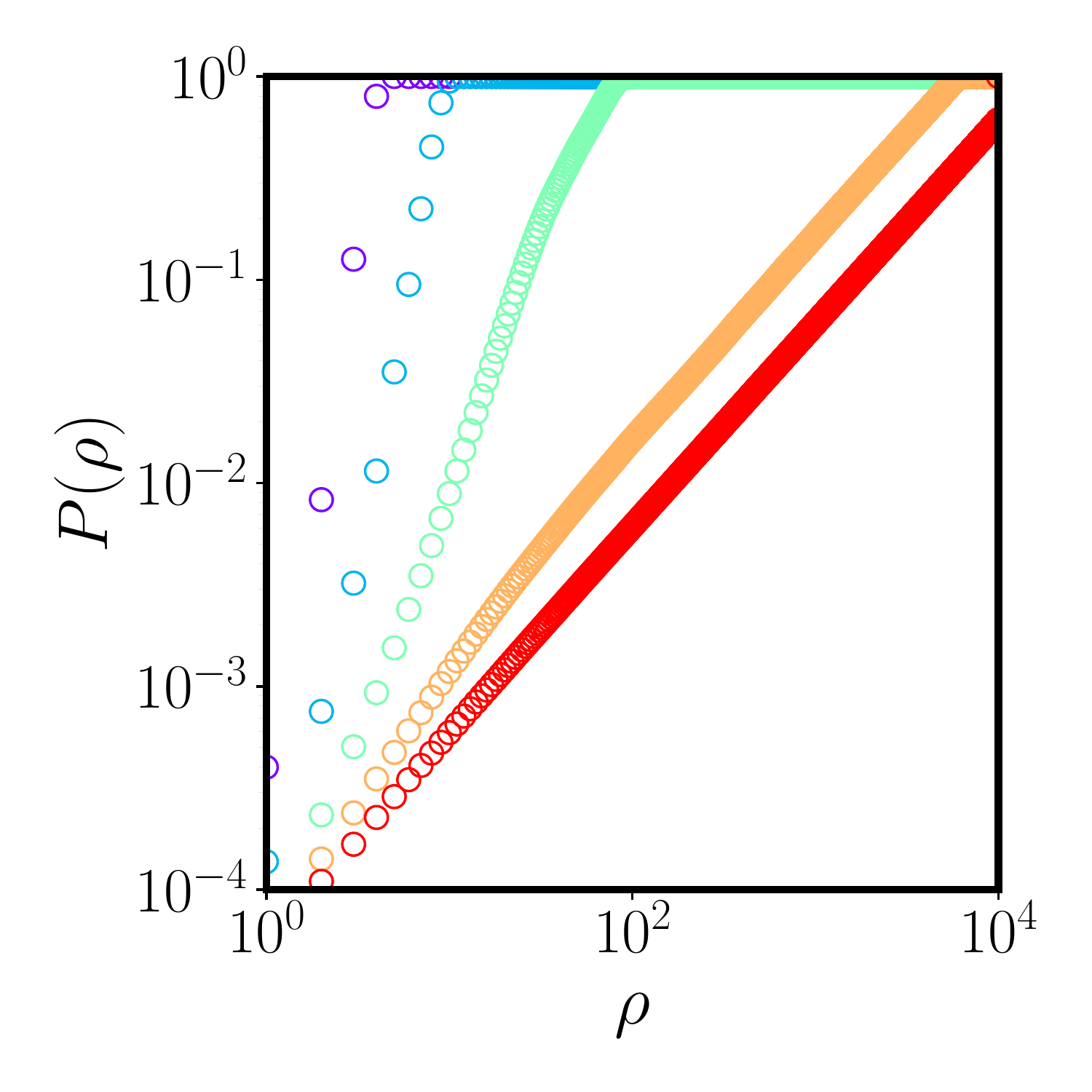}}
	\hfill
	\subfigure[]{\label{Fig2c}\includegraphics[width=.32\textwidth]{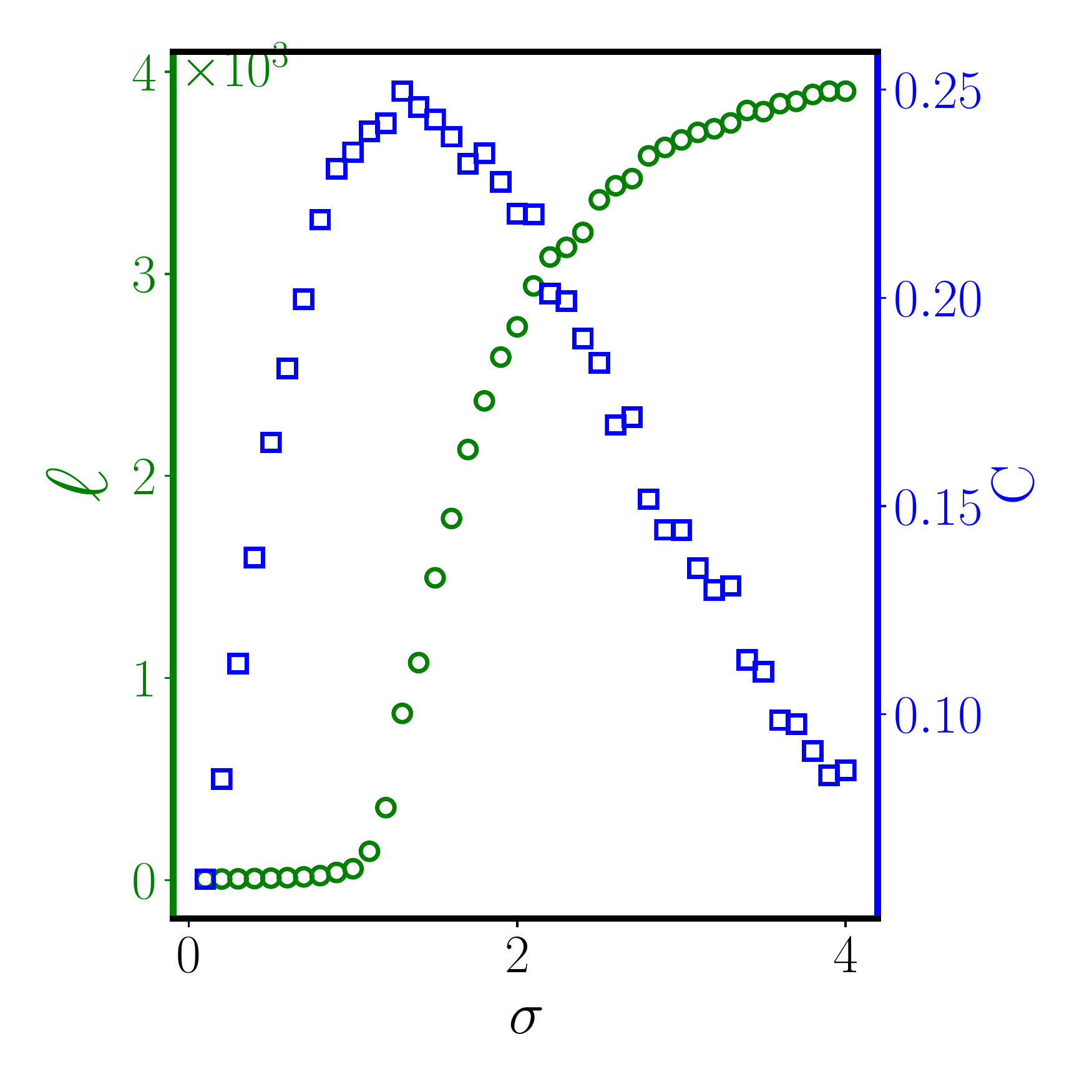}}

	\caption{Network statistics as function of $\sigma$ for $N = 2^{16}$ for the symmetric model with linear distance definition $\text{LRRR}^{(L)}$. (a)\,Degree distribution $p(k,\sigma)$ for five representative values of $\sigma=\{0.1, 0.5, 1.0, 1.5, 2.0\}$ from bottom to top (see legend). 
(b)\,Average fraction of nodes at a given topological distance $\rho$, $P(\rho)$, same $\sigma$ values (from top to bottom).
 (c)\,Mean minimum path $\ell)$ (green circles, left $y$-axis) and transitivity $C$ (blue squares, right $y$-axis).
 \label{Fig2}}
\end{figure*}

The degree of each node measures its number of neighbors, $k_i = \sum_{j=1}^N a_{ij}$. 
In the infinite-size limit $N\to \infty$, the degree distribution of the model is well approximated by a normal distribution as sketched in Fig.\,\ref{Fig2a} (see Appendix \,\ref{AppB} for more details). The mean of the distribution is $\kappa = 2 \zeta(\sigma+1)$ and its standard deviation is $\sigma_\kappa = \left[2\left( \zeta(\sigma+1) - \zeta(2\sigma+2)\right)\right]^{1/2}$, where $\zeta(s)$ is the Riemann zeta function. Notice that in the $\sigma\to 0$ limit both $\kappa$ and $\sigma_\kappa$ diverge, as $\lim_{s\to1}\zeta(s) =+\infty$. In the opposite limit $\sigma\to\infty$, the network converges to a ring chain with $\kappa= 2$ and $\sigma_{k}= 0$.

Many real-world networks are characterized by the presence of efficient pathways of communication. They can be quantified by the average path length $\ell$, which measures the mean topological distance between every pair of nodes over the network shortest paths: 
\begin{equation}
\ell = \frac{1}{N(N-1)} \sum_{i=1}^N \sum_{j\neq i} \rho_{ij},
\end{equation}
where $\rho_{ij}$ is the minimum number of links connecting nodes $i$ and $j$, i.e. the graph distance, which we also refer to as chemical distance. 
The cumulative distribution of $\rho_{ij}$, $P(\rho)$, indicates the average fraction of nodes that are within a radius $\rho$ of any given node. This  is shown in Fig.\,\ref{Fig2b} for different values of $\sigma$; it indicates how for small $\sigma$ the fraction of neighbors grows quickly with the distance $\rho$, whereas for $\sigma \gg 1$ the size of the nodes' neighbourhoods scale as a power-law of the distance -- the exponent of which gives the fractal dimension.

In general, low values of $\ell$ relative to the network size indicate the emergence of the small-world phenomenon\,\cite{watts1998collective}, associated with an efficient behavior of a communication network\,\cite{latora2001efficient}. 
More formally, a network is said to display such a property if $\ell$ grows proportionally to the logarithm of its size (i.e. number of  nodes)\,\cite{newman2011networks}, a feature of multiple graph models including Erd\H{o}s-R\'enyi networks\,\cite{erdHos1968random}.

An empirical property of many real-world networks is the presence of dense local structures, which can be for instance quantified by means of the network transitivity $T$, defined as 
\begin{equation}
T = \frac{\text{number of closed triangles}}{\text{number of open triads}},
\end{equation}
This feature is absent in Erd\H{o}s-R\'enyi and similar random graph models, but present in the LRRR model. 

Similarly to the original Watts-Strogatz model~\cite{watts1998collective}, the LRRR model is characterized by a regime of intermediate values of $\sigma$ which maximizes transitivity whereas displaying efficient communication structure, as shown in Fig.\,\ref{Fig2c}. The analysis pursued in Ref.\,\cite{biskup2004} already showed that the connectivity properties of this kind of models change their nature as a function of $\sigma$. The $\sigma>1$ case shall not possess small world properties, while displaying high clustering features. 
When the probability of long-distance connections grows for $\sigma<1$, the topology of the network changes and the topological distance seems to display sub-power-law scaling still maintaining finite clustering. The sub-power-law scaling of the topological distance in the thermodynamic limit can be proven exactly (see Ref.\,\cite{biskup2004}).
Finally, for $\sigma<0$ the network actually becomes small-world and the clustering vanishes in the thermodynamic limit. The transitions between these different topological regimes appear to be continuous similarly to conventional second-order phase transitions.

In the following we are going to show how these ``continuous transitions'' also influence the spectral properties of the network, even if the evolution of the spectrum appears to be far more involved than the one of the topological properties. 

\section{Spectral Properties}
\label{spec}
In order to evaluate the spectral dimension $d_{s}$ of the LRRR model, we consider the graph Laplacian $\mathcal{L}$ \cite{van2010graph},
\begin{equation}\label{graphlap}
\mathcal{L}_{ij}=\begin{cases}
   1 & \text{when $i=j$}, \\
   -\sqrt{\frac{1}{k_i k_j}} & \text{if $a_{ij}=1$},\\
   0 & \text{otherwise}
  \end{cases}.
\end{equation}and numerically evaluate its spectrum as a function of $\sigma$ for several realizations of the LRRR model. The convergence properties of the spectrum have been studied by calculating it for increasing network sizes up to $N=
2^{12}$. The numerical estimates for the spectrum upon increasing the number of network realizations or the network size have been shown to converge to the same function, indicating self-averaging properties and yielding a unique definition of spectral dimension $d_{s}$ in the thermodynamic limit. The numerical spectra of the LRRR model with linear distance definition ($\text{LRRR}^{(L)}$) have been compared with the ones obtained with the circular distance definition ($\text{LRRR}^{(C)}$), proving the isospectral property of the two models in the thermodynamic limit $N\to\infty$.
\begin{figure*}[ht!]
	\centering
	\subfigure[\hspace*{-2em}
	]{\label{Fig3a}\includegraphics[width=.32\textwidth]{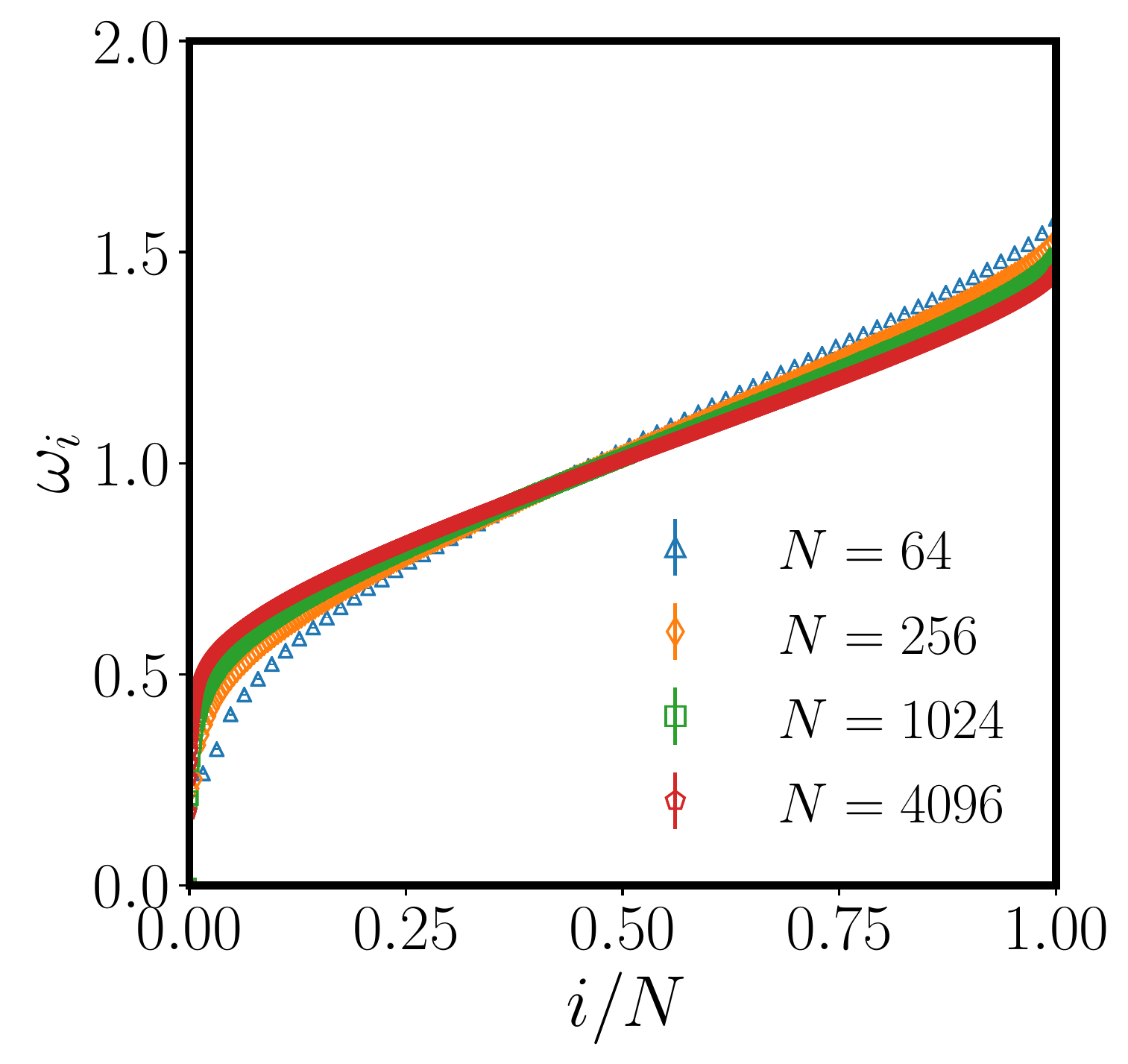}}
	\hfill
	\subfigure[\hspace*{-2em}
	]{\label{Fig3b}\includegraphics[width=.32\textwidth]{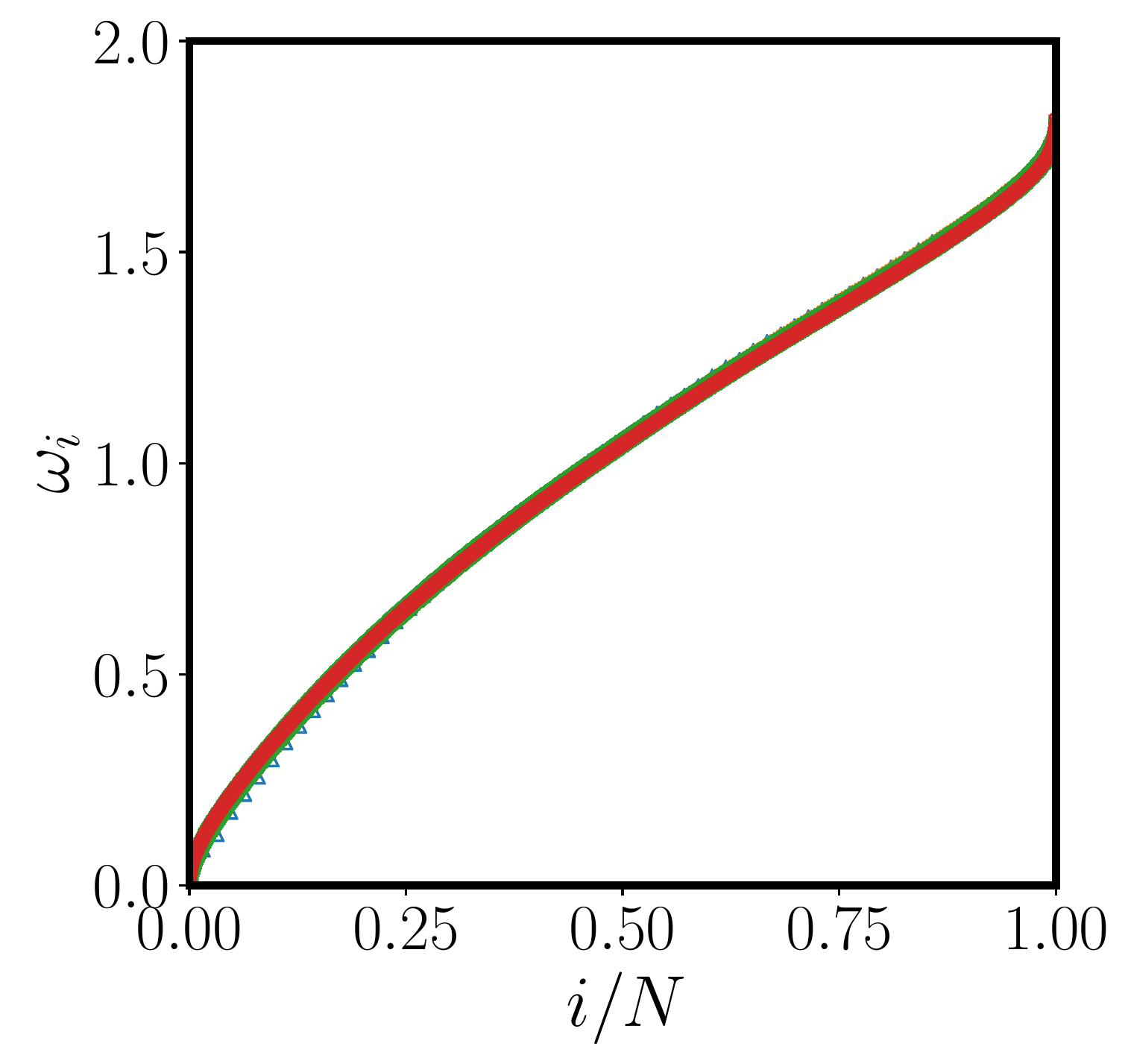}}
	\hfill
	\subfigure[\hspace*{-2em}
	]{\label{Fig3c}\includegraphics[width=.32\textwidth]{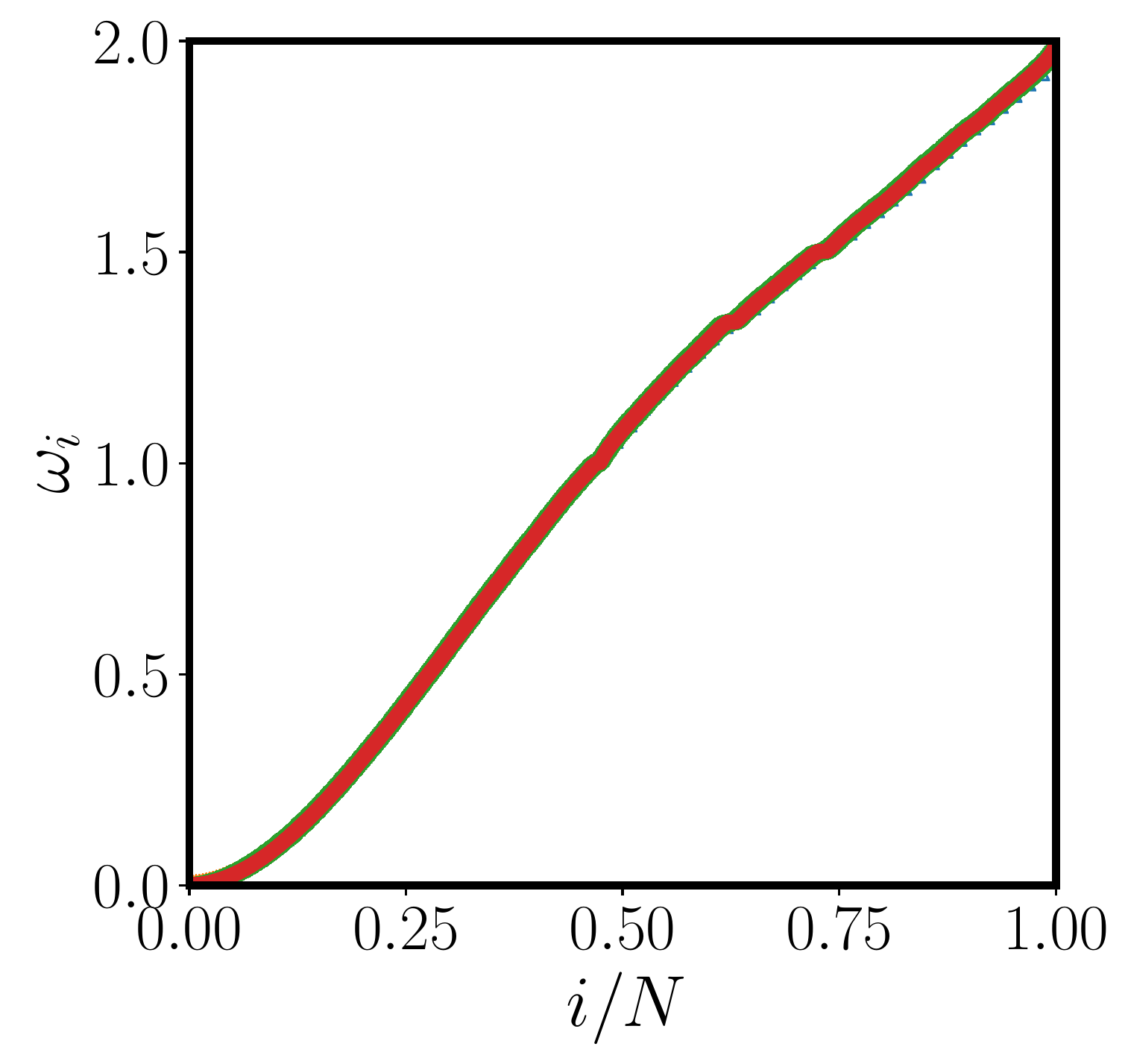}}
	\caption{Spectra (averaged over different realizations) for (a) $\sigma=0$, (b) $\sigma=0.5$, and (c) $\sigma=1.5$ for increasing system size ($N=64,\,256\,1024\,4096$ as, respectively, triangles, diamonds, squares, and circles). All three spectra refer to the linear model. In the intermediate decay range $\sigma>1$ flat regions appear in the energy spectrum, as can be seen in (c). These flat regions are caused by the appearance of localized eigenstates due to the presence of disorder. Since such localized states appear at high energy they do not influence low-energy properties such as the spectral dimension.
 \label{Fig3}}
\end{figure*}

The eigenvalues of the normalized Laplacian have been ordered based on their magnitude and are denoted by $\omega_{i}$, with $\omega_{1}=0$ being the eigenvalue of the steady-state
eigenvector. The ordered spectra are shown in Fig.\,\ref{Fig3} for $\sigma=0.0,0.5,1.5$ for various network sizes $N$. In all cases, the spectra converge to a well-defined functional form at large $N$, but finite-size corrections are more relevant for smaller $\sigma$. As expected, both the $\sigma>0$ cases present a continuous power-law behavior at $\omega_{i}\simeq 0$, indicating a low-energy density of states (DOS) for vibrational modes of the form
\begin{align}
\mathcal{D}(\omega)\propto \omega^{d_{s}-1}
\end{align}
with a finite value of the spectral dimension $d_{s}$. For $\sigma=0$ the spectrum appears to develop a finite gap $\omega_{2}-\omega_{1}\neq 0$ indicating that $d_{s}=\infty$. According to this analysis the point $\sigma=0$ does not only delimit the topological transition from a non small-world network $\sigma>0$ to a small-world one at $\sigma<0$, but also the appearance of a spectral gap in the model, which persists for all $\sigma<0$. 

In order to justify these observations on theoretical grounds one may construct the following analytically solvable model, which shares several features with the LRRR.  We consider the average over all possible realizations of the adjacency matrix of our model $\bar{a}_{ij}=p_{ij}=1/r_{ij}^{1+\sigma}$, which describes a fully connected weighted graph. The lack of translational invariance in the LRRR model is removed by the averaging procedure and the spectral dimension of the resulting graph is analytically known as
\begin{align}
\label{an_ds}
d_{s}=\begin{cases}
2/\sigma\,\,&\mathrm{if}\,\,0<\sigma<2\\
1\,\,&\mathrm{if}\,\,\sigma\geq2,
\end{cases}
\end{align}
(see Ref.\,\cite{Burioni1997}). 
In principle, we do not expect the estimate in Eq.\,\eqref{an_ds} to exactly reproduce the spectral dimension of the LRRR model, since taking the average directly on the adjacency matrix is not the same as taking it on the spectrum\,\footnote{Note that this procedure would correspond to take the \emph{annealed} version of the model in the language of disordered systems. Our study will be rather devoted to the \emph{quenched} case.}. 
However, based on the analogy with the problem of critical long-range  percolation, one may expect this result to be accurate both at $\sigma>2$, where the effect of long-range connectivity becomes irrelevant to the universal behavior, and at $\sigma<1/3$, where the universal behavior of the critical  percolation model lies in the mean-field regime\,\cite{Grassberger2013SIR, Grassberger20132DSIR,Gori2017}.

\begin{figure}[t!]
\centering
\includegraphics[width=.5\textwidth]{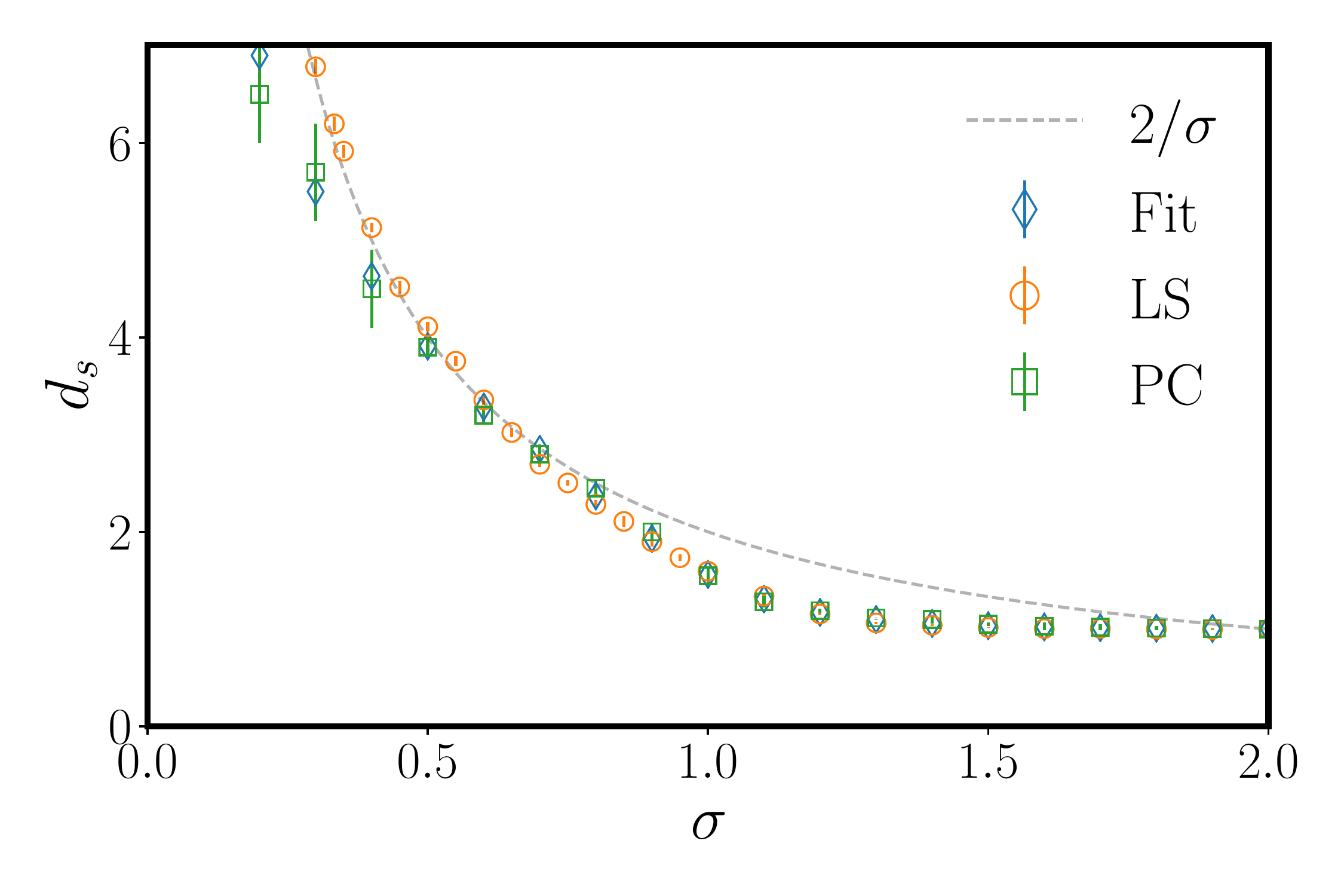}
\caption{The spectral dimension $d_{s}$ of the network model defined by Eq.\,\eqref{graphlap} as obtained by the finite-size scaling of the Laplacian spectrum (LS, orange circles), by the power-law return probability of the random walk (Fit, blue diamonds) and by the collapse of the return probability (PC, green squares). The dashed gray line represents the analytical expectation obtained by an approximate annealed model [see  Eq.\,\eqref{an_ds} and the discussion above]. Based on the discussion of Sec.\,\ref{model} these results correspond to the spectral dimension of a percolation cluster well in the supercritical regime.}\label{Fig4}
\end{figure}

A direct fit to the low-energy tails of the spectra in Fig.\,\ref{Fig3} does not yield reliable estimates for the spectral dimension values. Then, we shall rely on finite-size scaling properties. Indeed, in order for the spectrum to display the expected power-law behavior in the thermodynamic limit, each finite-size eigenvalue should exhibit the leading-order scaling
\begin{align}
\label{fs_eig}
\omega_{i}^{(N)}\propto N^{-2/d_{s}}.
\end{align}
Using Eq.\,\eqref{fs_eig} we can extract the spectral dimension from the finite-size scaling of the low lying eigenvalues (see Appendix \,\ref{LLS}). The resulting values for the spectral dimension as a function of $\sigma$ are reported as orange circles in Fig.\,\ref{Fig4}.  

\section{Random Walk} 
\label{RW}
\subsection{Return probability}
The spectral dimension controls both the scaling of the spectrum and the RW return probabilities, as mathematically argued in Ref.\,\cite{Burioni1996}. Then, if a particle is initially at the origin at time $t = 0$, the probability of finding it there at time $t$ is given by return probability $P_{0}(t)$\,\cite{Alexander1982, rammal1984random, Burioni1996, noh2004random}:
\begin{equation}\label{eq:p0t}
P_0(t) \sim t^{ -d_s/2},\quad t\gg1.
\end{equation}

In order to prove that such a universal relation is obeyed in our model, we numerically computed the return probability $P_0(t)$ on different realizations of our network. Initially the walker is placed on a random node $i$, and at each time step it jumps with uniform probability $1/k_i$ to a neighboring node. The walker is left to diffuse for a number of steps $\tau$ large enough to explore a macroscopic portion of the network. The results for the return probability shown in the paper have been obtained by averaging over $N_{R}=10^{5}$   random walk's trajectories on each network realisation with $\tau=10^{5},10^{6}$.

The value of $d_s$ as a function of $\sigma$ has been estimated using a maximum likelihood algorithm\,\cite{clauset2009power,klaus2011statistical}. For each value of the network size $N$ and of the decay exponent $\sigma$ this technique  requires the identification of an initial time $t_{min}$ and a final time $t_{max}$, between which one has to pursue the power-law fit. 
Indeed, the scaling behavior cannot appear at small times, as the return probability in this limit is highly influenced by the local structure of the LRRR model and by the absence of self-links, such that $P_0(2t-1)<P_0(2t)$ and then Eq.\,\eqref{eq:p0t} shall not be obeyed at small $t$. 
On the other end, finite-size effects still appear at very large times, particularly for small $\sigma$ as, due to the high connectivity, the random walkers can loop over the network faster in this case. The observation of more prominent finite-size effects at small $\sigma$ is consistent with the behavior observed in Fig.\,\ref{Fig3} for the Laplacian spectrum. 

Therefore, the scaling behavior of Eq.\,\eqref{eq:p0t} can only be observed for intermediate values of times and very large network sizes, leading to the necessity of identifying a proper time window to estimate the power-law decay exponent. In order to proceed with the $d_{s}$ estimations in this case, we select an initial sensible value of  $t_{max}$ for each $N$ and $\sigma$ pair and then optimize both the time boundaries $t_{min}$ and $t_{max}$ making use of a maximum likelihood algorithm adapted from Ref.\,\cite{alstott2014powerlaw}. 
Note that for small $\sigma$ ($\sigma<0.5$) the finite-size effects can appear before $t_{min}$, leading to the underestimation of $d_s$. Consequently, very large systems are necessary to estimate $d_s$ in this case (blue diamonds in Fig.\,\ref{Fig4}). 

\subsection{Finite-size effects on $d_s$}
\begin{figure}
\centering 
\includegraphics[width=.5\textwidth]{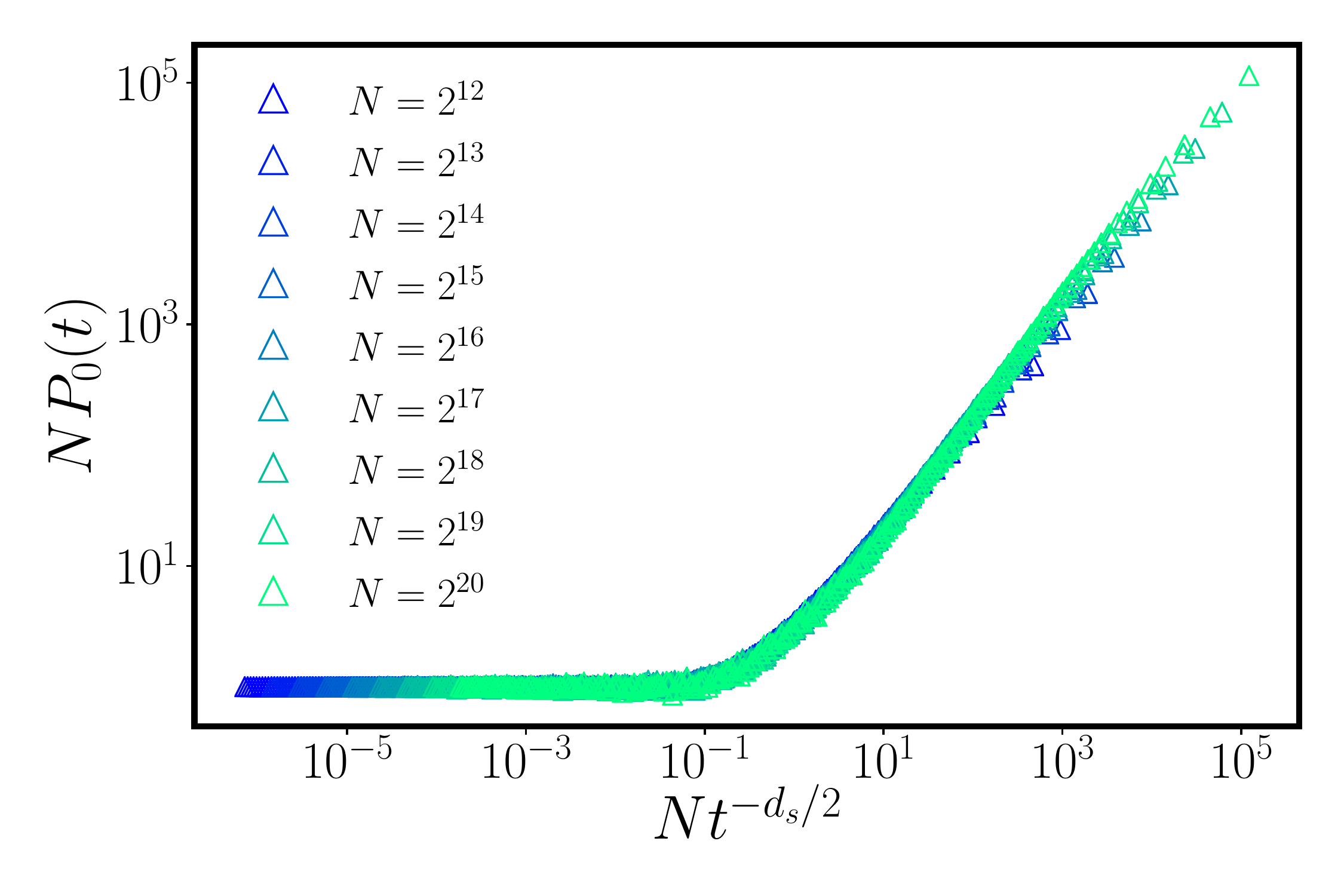}
\caption{Collapse of $P_0(t)$ for $\sigma=0.5$ and $N=2^i$, $i=12,13,...,20$ using the scaling function in Eq.\,\eqref{eq:collapse} (starting point from left to right). Both axes are in logarithmic scale. 
\label{Fig5} }
\end{figure}
Given the picture above, it is evident that finite-size corrections are expected to hinder the accuracy of the $d_{s}$ estimations from the random walk return probabilities, especially in the $\sigma \to 0$ limit where such corrections appear already at short times even for large sizes. In order to overcome these difficulties, we exploited the universal nature of the return probability and introduced the finite-size scaling of $P_0(t)$ as 
\begin{equation}
\label{eq:collapse}
P_0^N(t) = \frac{1}{N} f \left(Nt^{-d_s/2}\right),
\end{equation}
with $f(x)$ such that $f(x)\propto x$ for $x\gg 1$ and $f(x)\propto O(1)$ for $x\ll 1$. The latter finite-size scaling ansatz can be used to scale the return probabilities curves of different network sizes $P_{0}^{N}(t)$ on each other, thus yielding an estimate of $d_{s}$ by the optimal value for the collapse. This procedure is exemplified in Fig.\,\ref{Fig5} for $\sigma=0.5$; the optimal value for $d_{s}$ found in this case is $d_{s}\simeq 3.91$.  
 
The spectral dimension results from the probability collapse (PC) are shown as green squares in Fig.\,\ref{Fig4}. Finite-size effects also affect the collapse results for the spectral dimension $d_{s}$ at $\sigma\lesssim 0.5$, but the error bar estimates are more reliable with this method, when compared to the simple large time fit. In general, the comparison between random walk estimates, both by power-law fits (Fit) and by the return probability collapse (PC), yield consistent estimates in the whole $\sigma$ range and almost perfectly reproduce the Laplacian spectrum (LS) results for $\sigma\gtrsim 1/2$ corresponding to $d_{s}\lesssim 4$. The agreement between the different approaches furnishes a precise estimate of the spectral dimension in the most relevant regimes for critical $O(n)$ models, which exhibit non-mean-field universal behavior for $d\equiv d_{s}<4$.

Moreover, this agreement is found also for different boundary conditions (see Appendix \ref{LLS}): this proves the universality of the random walk scaling dynamics on this network model and provides a first hint of the universal role of the spectral dimension in this class of networks\,\cite{Burioni1996, Wu1995}. From the perspective of critical phenomena the LRRR network model described in Sec.\,\ref{model} corresponds to a long-range percolation cluster in the supercritical regime and, in principle, universality should not be expected\,\cite{Grassberger2013SIR, Grassberger20132DSIR,Gori2017}. Nevertheless, based on our numerical observations, the spectral dimension in this supercritical percolation cluster displays all the features of a universal quantity and also features a finite correction caused by disorder, which cannot be captured by the annealed model [see Eq.\,\eqref{an_ds} and the discussion above].
 
\begin{figure*}[ht!]
	\centering
	\subfigure[\hspace*{-2.5em}
	]{\label{Fig8a}\includegraphics[width=.32\textwidth]{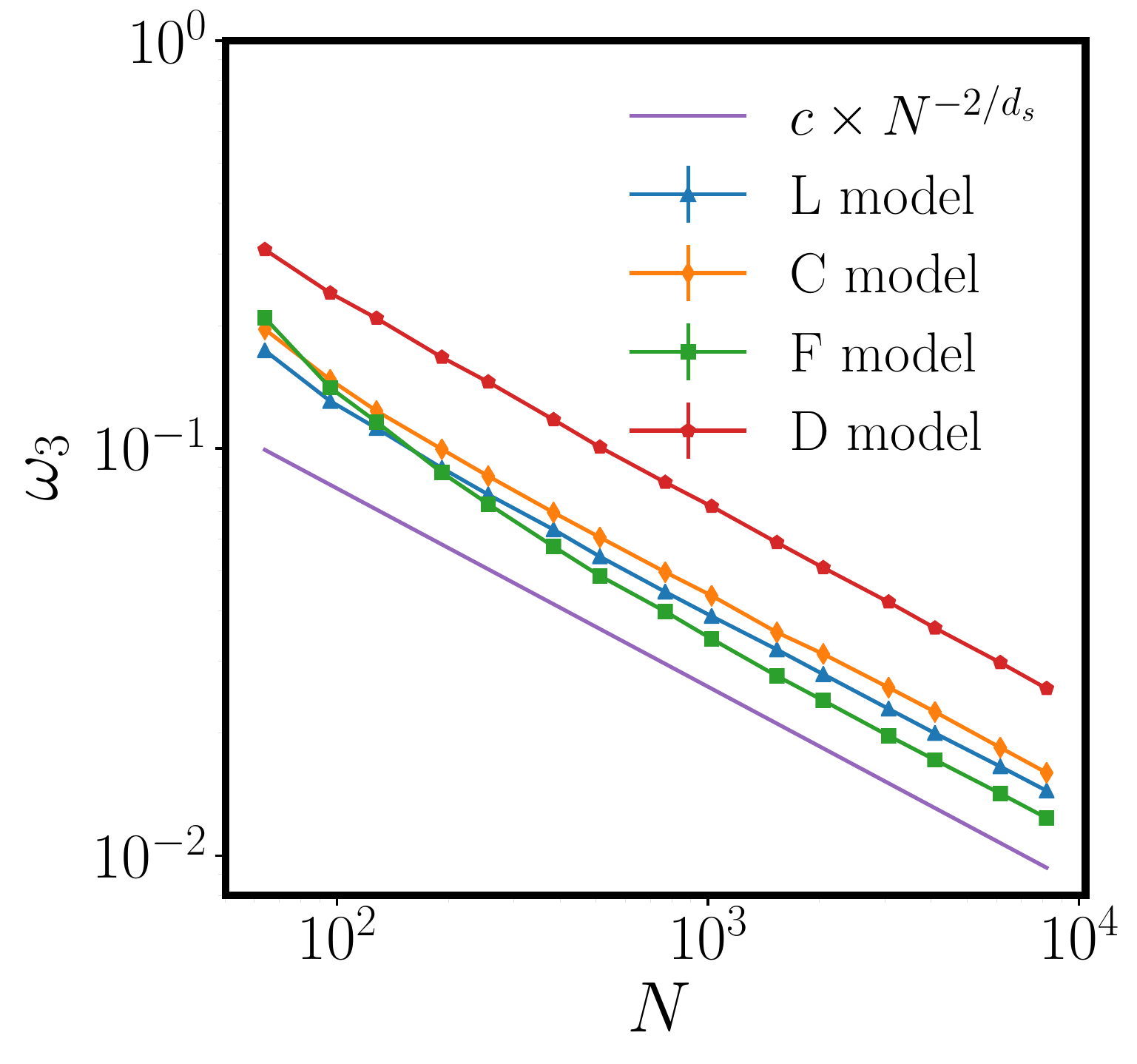}}
	\hfill
	\subfigure[\hspace*{-2.5em}
	]{\label{Fig8b}\includegraphics[width=.32\textwidth]{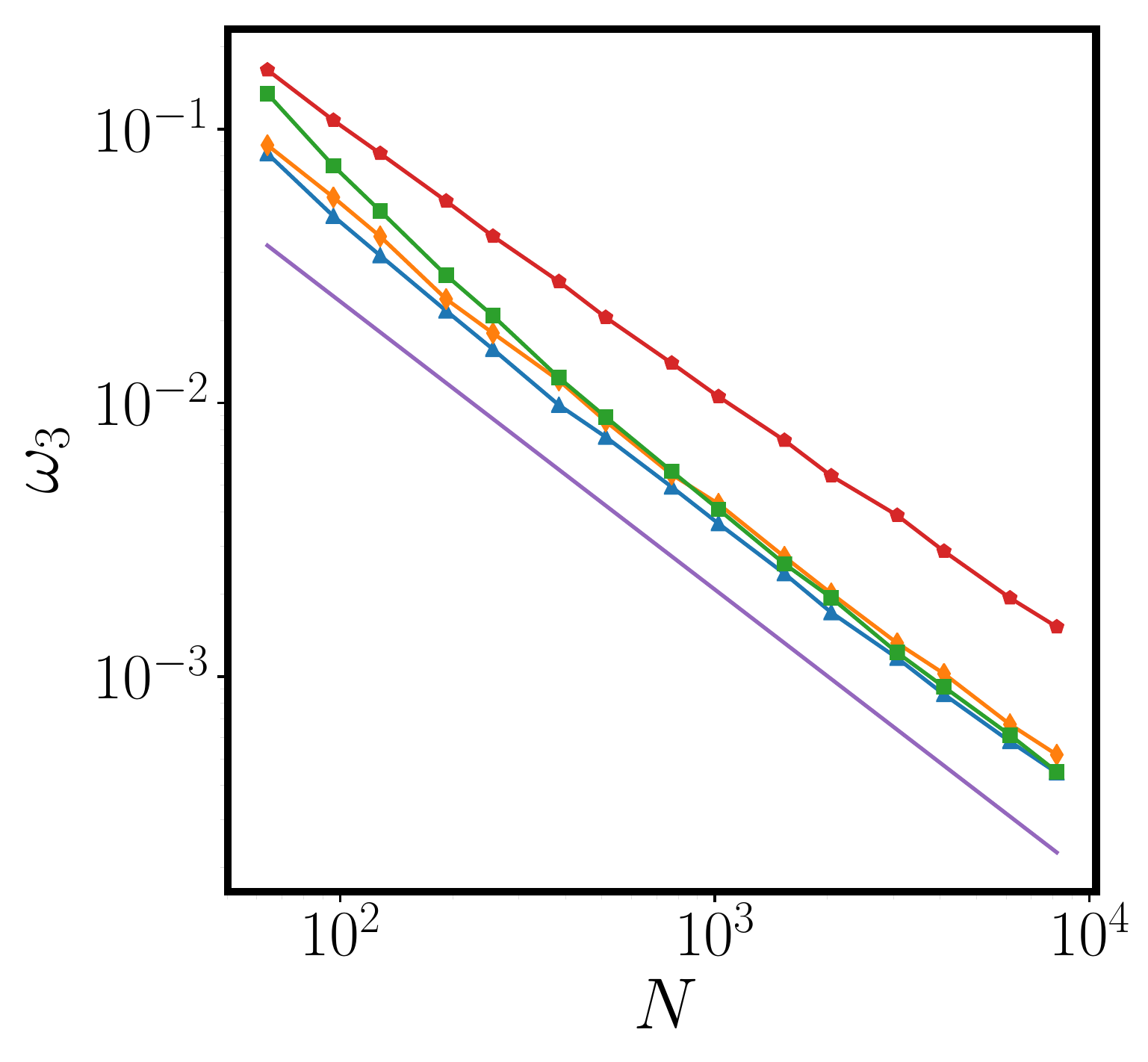}}
	\hfill
	\subfigure[\hspace*{-2.5em}
	]{\label{Fig8c}\includegraphics[width=.32\textwidth]{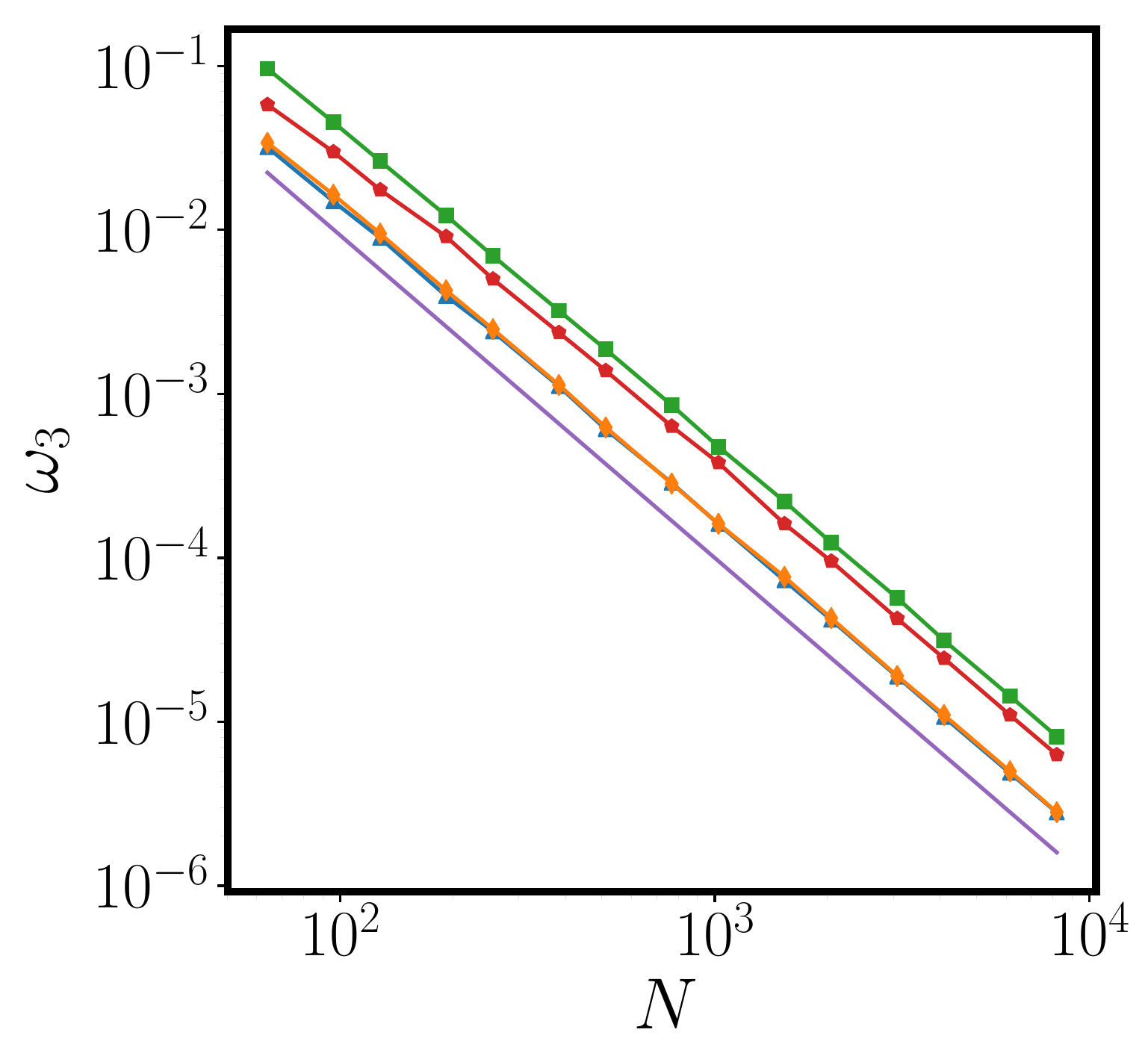}}
	\caption{Finite-size scaling of the third highest eigenvalue ($\omega_3$) of the Laplacian spectrum for the various definitions of the present network model: the linear ($L$) and circular ($C$) models, where the two distance definitions in Eqs.\,\eqref{l_def} and\,\eqref{c_def} have been employed (blue triangles and orange diamonds), the model with a thick backbone, i.e., both nearest-, next-nearest- and third-nearest-neighbors bonds active with unit probability (F model, green squares) and the D model, where the activation probability in Eq.\,\eqref{eq:netw_def} is doubled (red pentagones). The three panels refer to the decay parameters (a) $\sigma=0.5$, (b) $\sigma=0.9$, and (c) $\sigma=1.5$. All models display the theoretical finite-size scaling predicted by the spectral dimension reported in Fig.\,\ref{Fig4} (purple solid line in all sub-plots). The same scaling has been verified also for others higher-energy eigenvalues.
 \label{Fig8}}
\end{figure*}

It is worth restating that the present model does not have any conventional critical point, but the spectral dimension results in Fig.\,\ref{Fig4} are universal in the sense that they are not altered by microscopic modifications of the model under study, such as the ones described in Appendix \,\ref{AppB}, i.e., a change in the definitions of the distance or the introduction of additional short-range bonds (see Fig.\,\ref{Fig8}).
 
\section{A universality playground}
\label{univ_pl}

\subsection{Previous results}

The current understanding of critical phenomena is rooted in the study of prototypical models, which, in spite of their simplicity, can produce accurate predictions for real physical systems thanks to the universality phenomenon. 
Following this path, for most of the experimentally observed critical behavior it has been possible to construct a continuous field-theory model, which reproduces the appropriate universal quantities without any information about the discrete nature of the microscopic variables and the lattice structure. A paradigmatic example of this procedure can be found in the characterization of the universality of spontaneous symmetry breaking via the homogeneous $O(n)$-symmetric models. These models describe a vector order parameter $\boldsymbol{\varphi}$ with $n$ components, whose ground state value may be either $O(n)$ symmetric $|\boldsymbol{\varphi}_{0}|=0$ or spontaneously broken $|\boldsymbol{\varphi}_{0}|\neq0$. 

The early picture for the universal behavior of $O(n)$ models was first obtained by perturbative RG\,\cite{Wilson1974, Brezin1976, Brezin1993, Moshe2003} and has since then been complemented with several real-space and variational results\,\cite{Efrati2014,Kleinert2001, Zinn-Justin1996}. 
More recently, functional RG approaches\,\cite{Polchinski1984,Wegner1973,Wetterich1993} have been able to reproduce and extend previous findings, yielding the full universal landscape for $O(n)$ field theories\,\cite{Codello2013, Codello2015, Defenu:2017el, DefenuJHEP, Yabunaka2017, Yabunaka2018, Defenu2020fate}. With these extensive investigations, $O(n)$ models have become the general tool for the understanding of universal behavior in critical phenomena.

In these systems the only relevant parameters regulating universal behavior are the symmetry index $n$  and the Euclidean spatial dimension $d$: they control the phase space for critical fluctuations by altering, respectively, the number of fluctuating modes and the low-energy tails of the density of states (DOS). Interestingly, the universal properties can be analytically continued to the two-dimensional plane $(d,n)\in\mathbb{R}^{2}$, leading to a complex phase diagram which has been a fundamental ingredient in the understanding of universality\,\cite{Cardy1980, Peled2017, Stanley1968, deGennes1972, Balian1973, Fisher1973}. 

\begin{figure}[ht!]
\centering
\includegraphics[width=.5\textwidth]{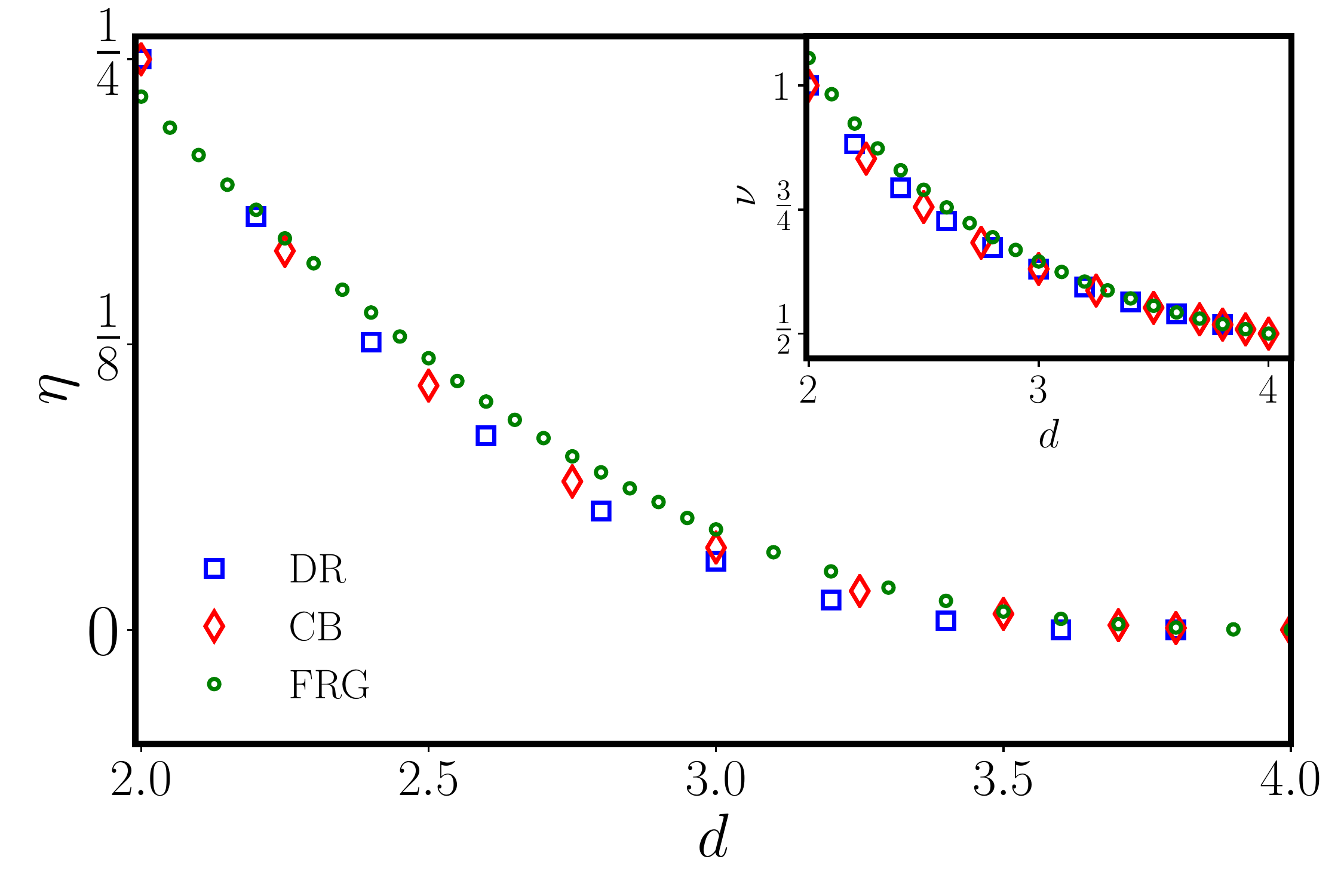}
\caption{Critical exponents $\eta$ and $\nu$ (inset) of the Ising model as a function of $d\in[2,4]$ from dimensional regularization (DR, blue squares)\,\cite{Holovatch1993}, conformal bootstrap (CB, red diamonds)\,\cite{ElShowk2012, ElShowk2014, Cappelli2019}, and functional renormalization group (FRG, green circles)\,\cite{Codello2013,Codello2015}. \label{Fig7}}
\end{figure}

The intricacies regarding the proper definition of dimension on graphs have, up to now, hindered the validation of the existing theoretical results for universal behavior in fractional dimension on discrete inhomogeneous structures. Yet, theoretical investigations alone have reached a fair degree of consistency and unity among each other, yielding a comprehensive picture of the critical exponents of $O(n)$ models in the continuum with Euclidean dimension $2\leq d \leq 4$\,\cite{Zinn-Justin1996,Pellissetto2002,Kleinert2001,Codello2013, Codello2015}. 
For integer Euclidean dimensions $d\in\mathbb{N}$, this picture can be verified by numerically exact results obtained by Monte Carlo (MC) simulations\,\cite{Pellissetto2002}, and, at least for the Ising model ($n=1$), conformal bootstrap results, which are believed to be exact and also extend to $d\in\mathbb{R}$\,\cite{ElShowk2012} (see Fig.\,\ref{Fig7}).

The results depicted in Fig.\,\ref{Fig7} prove the capability of current theoretical approaches to provide reliable estimates of universal quantities in $O(n)$ field theories. Yet, no numerical confirmation or exact proof of the applicability of these results to microscopic discrete models exists. As anticipated above, the natural candidate for the dimension, as a relevant parameter for universality, on graphs and complex networks is the \emph{spectral dimension}\,\cite{Cassi1992, Burioni1996, Burioni1999}. Indeed, the existence of the critical point for $O(n)$-symmetric models on complex networks is solely determined by the value of the spectral dimension, at least as long as $d_{s}>2$\,\cite{Cassi1996, Burioni1999, Burioni1999jpa, Bradde2010}.  Finally, the universal properties of most exactly solvable models, including the $O(n)$ models in the $n\to\infty$ limit, only depend on $d_{s}$\,\cite{Hattori1987, Cassi1999, Burioni2000, Buonsante2000}. 

The numerical confirmation of the above picture in correlated critical models is lacking, even in the simpler case of continuous symmetry $n\geq 2$ where no universal behavior is found at $d_{s}\leq2$.  This is mostly due to the difficulty of identifying proper graph models that present both a tuneable spectral dimension and a stable numerical behavior in the limit of large size. Indeed, mathematically exact derivations of the spectral dimensions of fractals are known only in few cases, usually with $d_{s}<2$\,\cite{Maritan1986, Thouy1995, Freiberg2010}, while numerical simulations need large sample sizes and long computation times\,\cite{Rudra1990}. 

In this work, we have shown that the long-time scaling of random walk dynamics on the LRRR network is solely determined by the spectral dimension and, therefore, may be regarded as universal. Based on the results of Ref.\,\cite{Burioni1996}, this scaling corresponds to that of a Gaussian field theory at the critical point and constitutes a first confirmation of the role of the spectral dimension as a control parameter for universal behavior. Moreover, based on the investigations of Refs.\,\cite{Grassberger2013SIR, Grassberger20132DSIR,Gori2017}, it is straightforward to infer that prototypical models such as percolation will display a non-trivial critical point when placed on the LRRR graph.  These two statements make the LRRR network model unique between the few already existing candidates of graphs with tuneable $d_{s}$, which are typically pathological, since they do not exhibit non-trivial random walk scaling\,\cite{Burioni1994fract} nor correlated critical points\,\cite{Wu1995}. 

\subsection{Universality of the spectral properties}

In order to relate our studies to the aforementioned picture of universality, it is useful to consider our findings in the perspective of the long-range percolation problem discussed in Refs.\,\cite{Grassberger2013SIR, Grassberger20132DSIR, Gori2017}. In this problem each possible link of a one-dimensional chain is present with probability $p_{ij}=p\,r_{ij}^{-(d+\sigma)}$. This leads to the existence of a percolation threshold $p_{c}$ at which critical scaling appears. Then, the complex network model introduced in Sec.\,\ref{model} can be regarded as the giant cluster of the long-range percolation problem well inside the percolating regime with $p=1\geq p_{c}\,\,\forall\,\sigma$.

From this point of view, the problem of \emph{long-range} percolation on a one-dimensional ring 
is equivalent to the one of \emph{nearest-neighbour} percolation on the complex graph analyzed in this paper. In Ref.\,\cite{Gori2017} the critical properties of the long-range problem have been related to those of nearest-neighbor percolation in an effective fractional dimension $d_{\mathrm{eff}}=\frac{2-\eta_{\mathrm{sr}}}{\sigma}d$, where $\eta_{\mathrm{sr}}$ is the anomalous dimension of the nearest-neighbour problem in dimension $d_{\mathrm{eff}}$, as usual in the long-range literature\,\cite{Angelini2014, Defenu2015, Defenu:2017dc, Gori2017}. The contribution of the anomalous dimension to the effective dimension is necessary to take into account the renormalization of the field scaling dimension as compared to the case of quadratic models.

\begin{figure}
\centering 
\includegraphics[width=.5\textwidth]{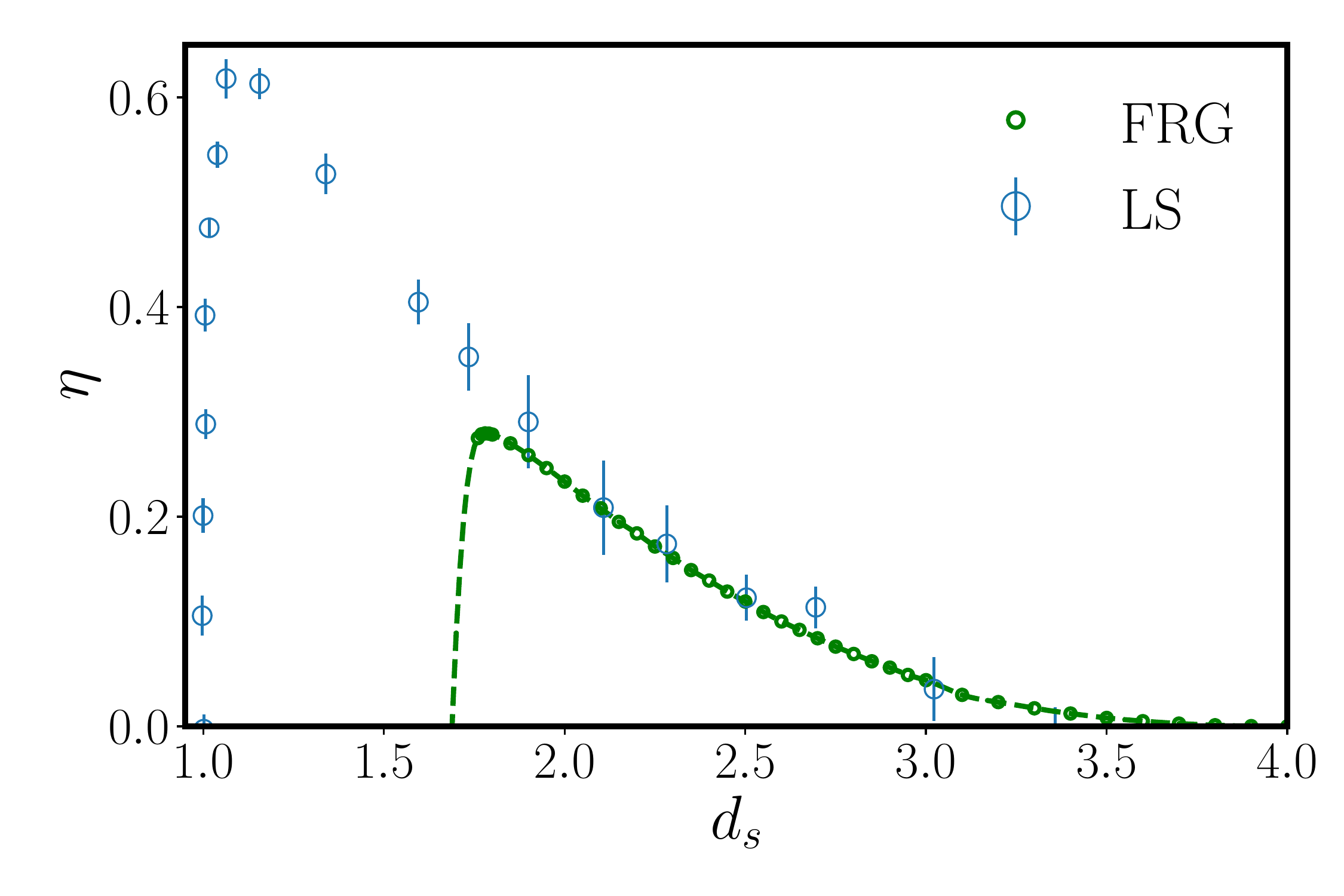}
\caption{The anomalous dimension $\eta$ generated by long-range disorder, as defined by Eq.\,\eqref{ren_ds}, is compared to the FRG estimates of the anomalous dimension of the Ising model as a function of the Euclidean dimension $d$.  The data are shown as large (blue) and small (green) circles, while the  dashed curve represents a numerical extrapolation of the FRG data.
\label{Fig7_new} }
\end{figure}

Therefore, in analogy with the case of long-range critical phenomena, one may interpret the deviation between the spectral dimension of the LRRR model and the annealed estimate \eqref{an_ds} as an \emph{anomalous dimension} $\eta$ due long-range disorder according to the formula
\begin{align}
\label{ren_ds}
d_{s}=\frac{2-\eta}{\sigma}\,.
\end{align}
In Fig.\,\ref{Fig7_new}, the values of $\eta$ obtained via Eq.\,\eqref{ren_ds} are reported and compared with those of the Ising model obtained by FRG, which also yields non-vanishing results for $d<2$, specifically down to $d\approx 1.76$. The similarity between the two curves is not surprising as the anomalous dimensions of several correlated models have similar trends\,\cite{Codello2013, Codello2015}.

The comparison between the anomalous dimension curves in Fig.\,\ref{Fig7_new} has to be considered with a grain of salt. Indeed, the anomalous dimension defined by Eq.\,\eqref{ren_ds} does not represent a ``canonical'' critical exponent, since it is not even defined at a critical point. It rather represents an anomalous contribution to the spectral dimension of the LRRR network coming from disorder, as it quantifies the discrepancy from the annealed (non-disordered) model. This quantity is universal in the sense that it is stable across different realizations of the network and it does not depend on the particular choice of distance and boundary conditions. Moreover, the quantity $\eta$ in Eq.\,\eqref{ren_ds} displays all the qualitative features of the anomalous dimension of a $\varphi^{4}$ theory, as proven by the comparison in Fig.\,\ref{Fig7_new}. The unexpected quantitative agreement between the quantity $\eta$ and the anomalous dimension of continuous scalar $\varphi^{4}$ theories furnishes further evidence of the intriguing scenario opened by the scaling properties of complex network structures with finite spectral dimension and of its connection with traditional studies in critical phenomena.

\section{Conclusions}
\label{concl}

From the human brain\,\cite{bullmore2009} to particles and grains \cite{papadopoulos2018network}, networks are the natural tool for a formal description of systems made up of many interacting agents. Over the years, a wide variety of dynamical processes have been studied on networks, from epidemic spreading\,\cite{pastor2015epidemic} and diffusion\,\cite{masuda2017random} to synchronization\,\cite{arenas2008synchronization}. While it is well known that the exact  form of interactions can affect the emergent dynamics\,\cite{boccaletti2006complex}, the link between network structure and critical behavior\,\cite{dorogovtsev2008critical} is still far from understood. 

Going beyond the individual assessment of specific network features, including average path length, clustering coefficient~\cite{watts1998collective} or the heterogeneity of the degree distribution \cite{barabasi1999emergence}, in this paper we turned our attention to a more fundamental definition of network dimension, i.e., the spectral dimensions, which may govern universality in interacting systems. To this end, we have introduced a complex network model based on the one-dimensional percolation problem studied in Refs.\,\cite{Grassberger2013SIR, Grassberger20132DSIR,Gori2017}. This model, which we name LRRR, coincides with a one-dimensional generalisation of the Kleinberg model\,\cite{Kleinberg2000}. 

Using extensive numerical simulation, we have characterized the spectral dimension $d_{s}$, which appears both in the scaling of the Laplacian spectrum and in the random walk return rates, and proved that it can be continuously tuned in the interval $d_{s}\in [1,\infty)$. Taken together, our model offers a valuable tool to study dynamical phenomena in  presence of a complex, but now well characterized, spectral landscape, offering insights into fundamental aspects of universal and critical behavior arising from network dynamics. 

Meanwhile, the network community has recently seen a surge of interest in the spectral dimension to connect the topological
and geometrical properties of a network~\cite{mulder2018network,boguna2020network} with its dynamics. So far, these explorations have mostly focused on systems interacting beyond traditional pairwise mechanisms~\cite{battiston2020networks}. In particular, the study of the spectral dimension of certain simplicial complexes~\cite{wu2015emergent,bianconi2016network} via a renormalization group approach has yielded accurate relations between $d_{s}$ and the topological dimension of the model\,\cite{bianconi2020spectral}.
Moreover, the spectral dimension was shown to be crucial to determine the synchronization properties of the simplicial implementation of the Kuramoto model recently suggested in \cite{millan2019synchronization}, and to affect diffusion
properties at long time scales \,\cite{millan2020explosive, torres2020simplicial}.  These works could only consider a finite number of $d_{s}$ values very close to the topological dimension of the building blocks of their network and they do not offer any realisation of tuneable $d_{s}>2$ values. We are convinced that the introduction of a model with continuously tuneable spectral dimension such as the LRRR model will pave the way to further investigations of the role of topology in network dynamics.

\textit{Acknowledgements}: The authors acknowledge fruitful discussions with Fabiana Cescatti, Miguel Ib\'a\~nez-Berganza and Andrea Trombettoni during various stages of this work. This work is supported by the Deutsche Forschungsgemeinschaft (DFG,
  German Research Foundation), Project-ID No. $273811115$ (SFB$1225$ ISOQUANT)
  and under Germany's Excellence Strategy EXC2181/$1-390900948$ (the   Heidelberg STRUCTURES Excellence Cluster). 
  F.B. acknowledges partial support from the ERC Synergy Grant No. $810115$ (DYNASNET).
A.P.M. also acknowledges support from the  ``European Cooperation in Science \& Technology'' (COST action Grant No.CA15109) and from ZonMw and the Dutch Epilepsy Foundation, Project No. $95105006$.

\appendix

\section{Low-lying spectrum}
\label{LLS}
\begin{figure}[ht!]
\centering
	\subfigure[]{\label{FigApp1a}\includegraphics[width=.45\textwidth]{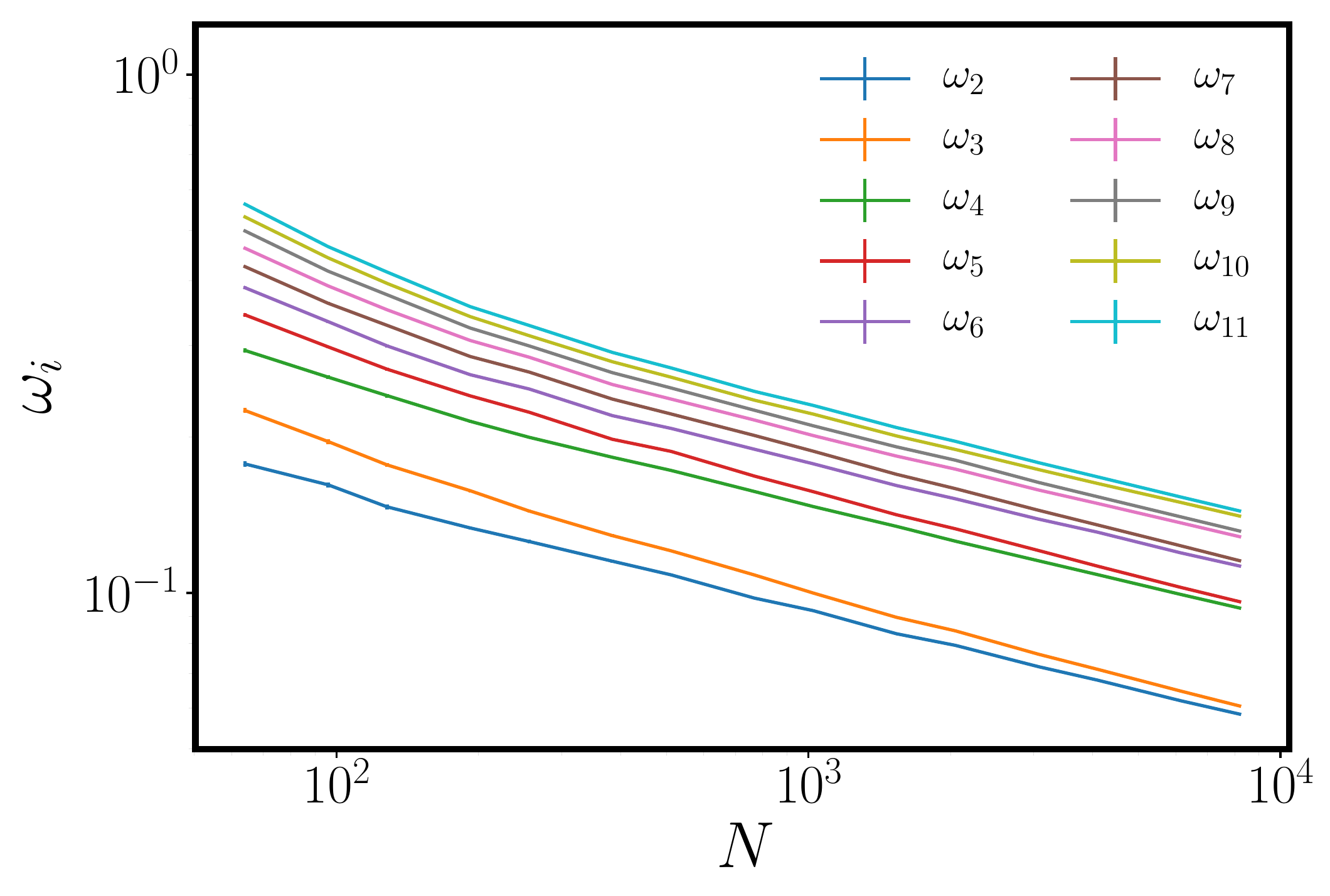}}
	\hfill
	\subfigure[]{\label{FigApp1b}\includegraphics[width=.45\textwidth]{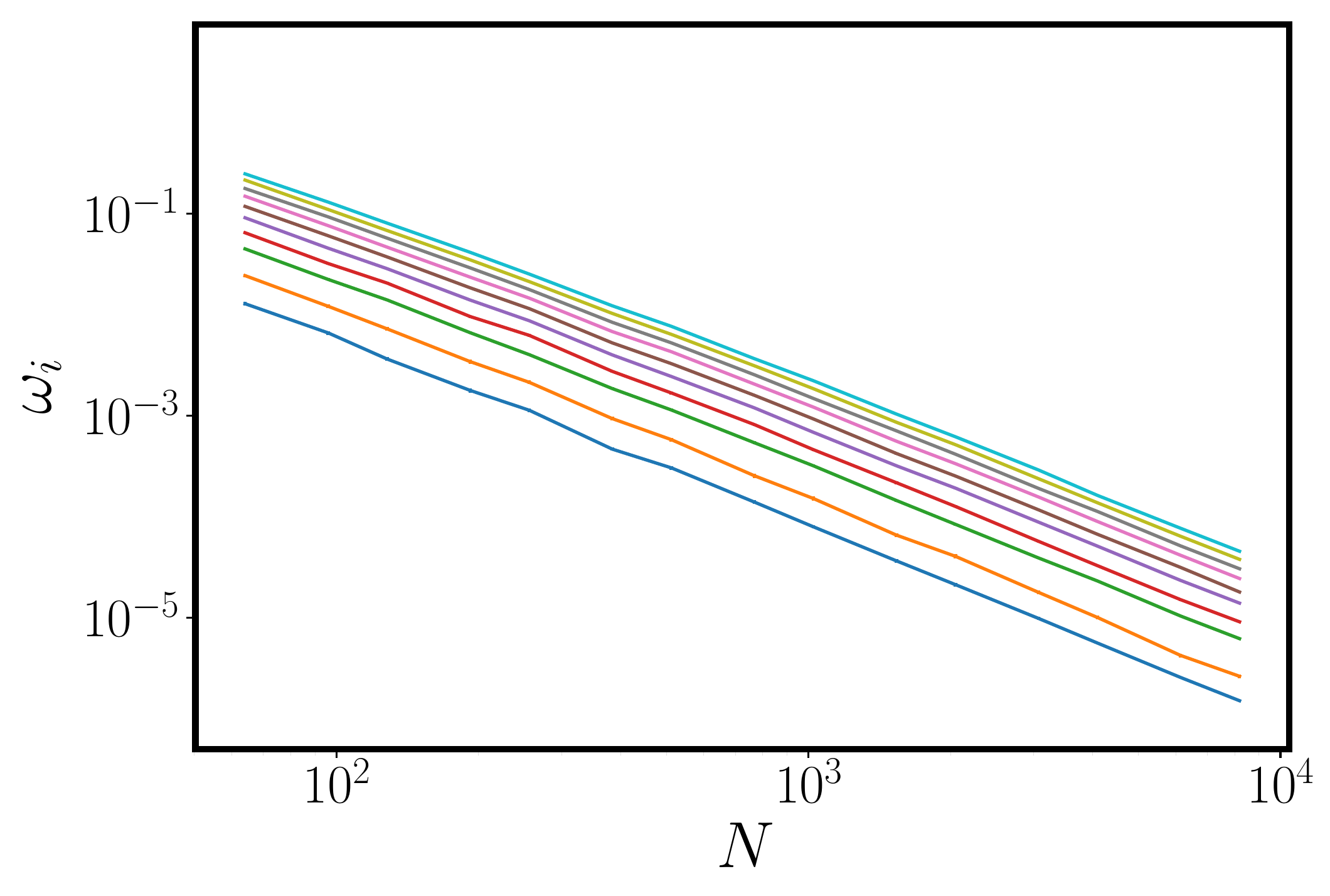}}
\caption{First $10$ nonzero eigenvalues $\omega_{i}$ with $i=2,3,\cdots,11$ from bottom to top for (a) $\sigma=0.5$ and (b) $\sigma=1.3$ as a function of the system size $N$ in the circular model.\label{FigApp1}}
\end{figure}

We detail here the structure of the low-lying spectrum as obtained by exact diagonalization of the graph Laplacian \eqref{graphlap}.
In Fig.\,\ref{FigApp1} the first $10$ nonzero eigenvalues are depicted as a function of the system size (averaged over $128$ realizations) for sizes up to $N=2^{13}$ for two different values of $\sigma$.
As one can notice the power-law decay is clearly attained for larger systems for all the depicted eigenvalues but the first eigenvalues display some oscillations that become smaller for the higher eigenvalues. 
Moreover, especially for small $\sigma$ the eigenvalues tend to organize into doublets, thus estimation of $d_s$ from a single eigenvalue could be affected by overshoot or undershoot.
In order to minimize the above effects, we obtained our best estimates from the average of $\omega_{10}$  and $\omega_{11}$ fitted with a power law $\propto N^{-2/d_s}$, for sufficiently big sizes $N\geq 2^{10}$.
The resulting estimations for the linear and circular models
were mutually compatible supporting our estimation.
Numerical work was not restricted  to linear and circular models,
but we also considered: graphs with a thicker backbone (with next-to-nearest and next-to-next-to-nearest neighbors always
turned on), which we refer as model $F$, models with non-backbone probabilities halved (not shown) ($p_{ij}\rightarrow p_{ij}/2$) and doubled ($p_{ij}\rightarrow 2 p_{ij}$), which we refer as model $D$.
All of these models, albeit possessing a different spectrum reflecting  the different non-universal (high-energy) features,
share the same low-lying spectrum behavior lending support to the universality of $d_s$ for models with the same decay exponent $\sigma$. 
In order to substantiate the present claim, we show the finite-size scaling for the third lowest eigenvalue for all the aforementioned models in Fig.\,\ref{Fig8}, the same result has been also verified for other higher eigenvalues.

\section{Computational method to estimate $d_s$ from $P_0(t)$}
\label{AppC}
\begin{figure}[htb!]
	\subfigure[]{\label{Fig:SIa}\includegraphics[width=.45\textwidth]{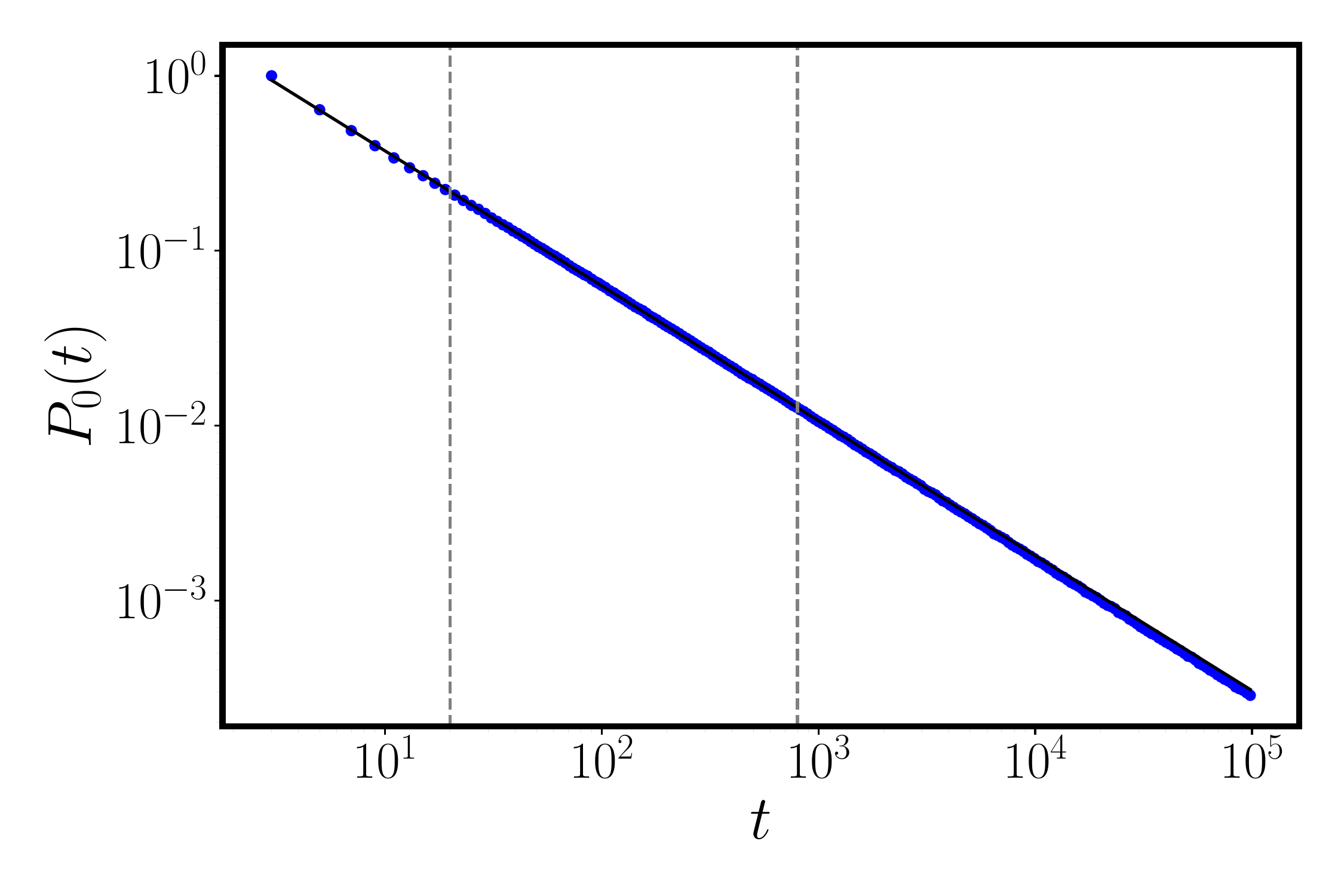}}
	\subfigure[]{\label{Fig:SIb}\includegraphics[width=.45\textwidth]{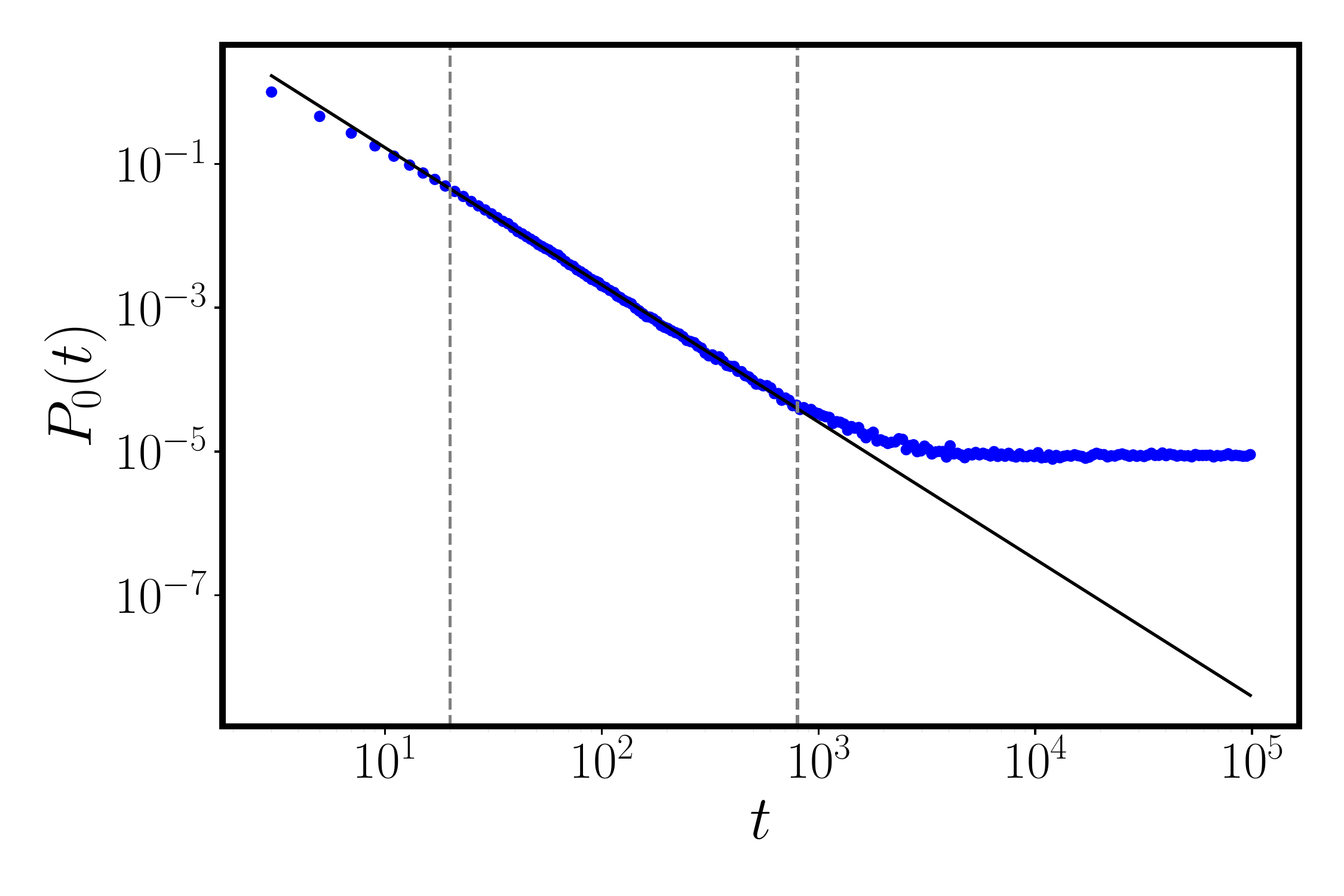}}
	\caption{Example of $d_s$ fit. $P_0(t)$ is fitted in the region between the dashed lines. In  \textbf{(a)}, $\sigma=1$, and finite-size effects are not strong in this range of $t$. Estimated value: $d_s = 1.569 \pm 0.004$. In  \textbf{(b)}, on the contrary, $\sigma=0.5$ and finite-size effects appear early on. Estimated $d_s = 3.79\pm 0.03$.  \label{fig:SI_dS}}
\end{figure}

An illustration of the method used to measure $d_s$, as indicated in the main text, is shown for two different values of $\sigma$ in Fig.\,\ref{fig:SI_dS}. First, $P_0(t)$ is represented in a log-log scale, and a power-law function via a maximum likelihood algorithm \cite{alstott2014powerlaw} that finds optimal values of $t_{min}$ and $t_{max}$. In case of a pronounced finite-size effect, as in Fig.\,\ref{Fig:SIb}, an initial $t_{max}$ is consider to avoid fitting of the flat part of $P_0(t)$.

\section{Characterization of LRRR networks}
\label{AppB}
\begin{figure}[!ht]
\includegraphics[width=\columnwidth]{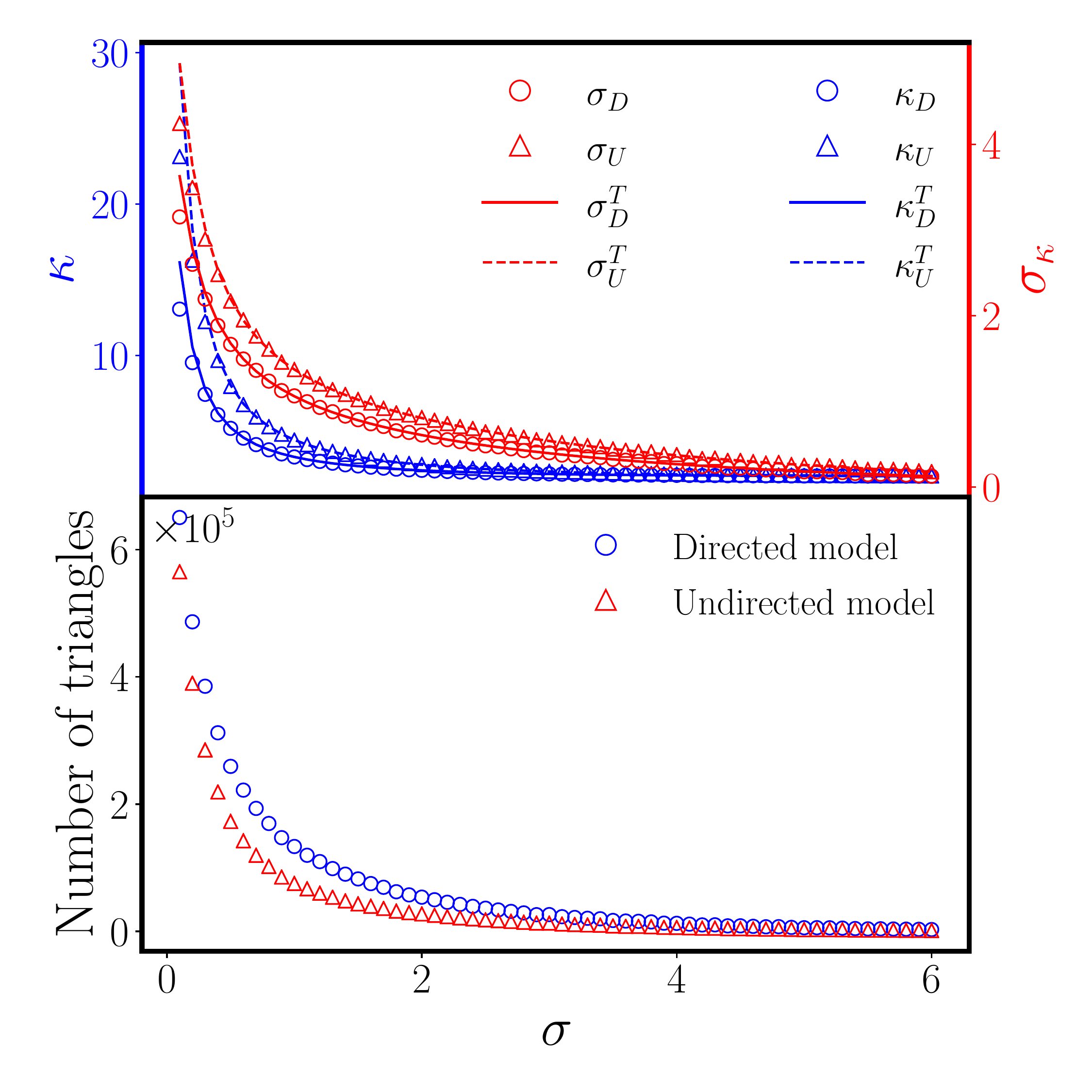}
\caption{(a) Dependence of the mean degree $\kappa$ (left $y$-axis) and its standard deviation $\sigma_\kappa$ (right $y$-axis) on $\sigma$, both for the directed and undirected cases. \textbf{(b)} Number of triangles in LRRR as a function of $\sigma$, both for the directed and undirected cases. \label{fig:SI1}}
\end{figure}

In the infinite-size limit ($N\to \infty$), the mean degree of the directed LRRR networks, without imposing network symmetry, is given by
\begin{equation}
\kappa_D^T = 2 \zeta(\sigma+1),
\end{equation}
whereas the standard deviation of the degrees is
\begin{equation}
\sigma_D^T = \left[2\left( \zeta(\sigma+1) - \zeta(2\sigma+2)\right)\right]^{1/2}.
\end{equation}
In our random walkers analysis, the networks are made symmetric by defining $a^S_{ij} = \max(a_{ij},a_{ji})$, that is, an undirected link is placed between nodes $i$ and $j$ when at least one directed edge is present. In this condition the mean degree and its standard deviation are given by:
\begin{align}
\kappa_U^T = 2\zeta(\sigma+1) - \zeta(2\sigma+2),
\end{align}
\begin{align}
\sigma_U^T &= [2( 2\zeta(\sigma+1) - 5\zeta(2\sigma+2) \nonumber\\
&+4\zeta(3\sigma+3) -\zeta(4\sigma+4) )]^{1/2}.
\end{align}
The theoretical curves are shown in Fig.\,\ref{fig:SI1}(a) together with the respective numerical estimates. 
These differences between the topological properties of the undirected and symmetrized directed LRRR networks do not influence the low-energy spectrum and, thus, do not alter the spectral dimension results.
\begin{figure}[!ht]
\centering
\includegraphics[width=\columnwidth]{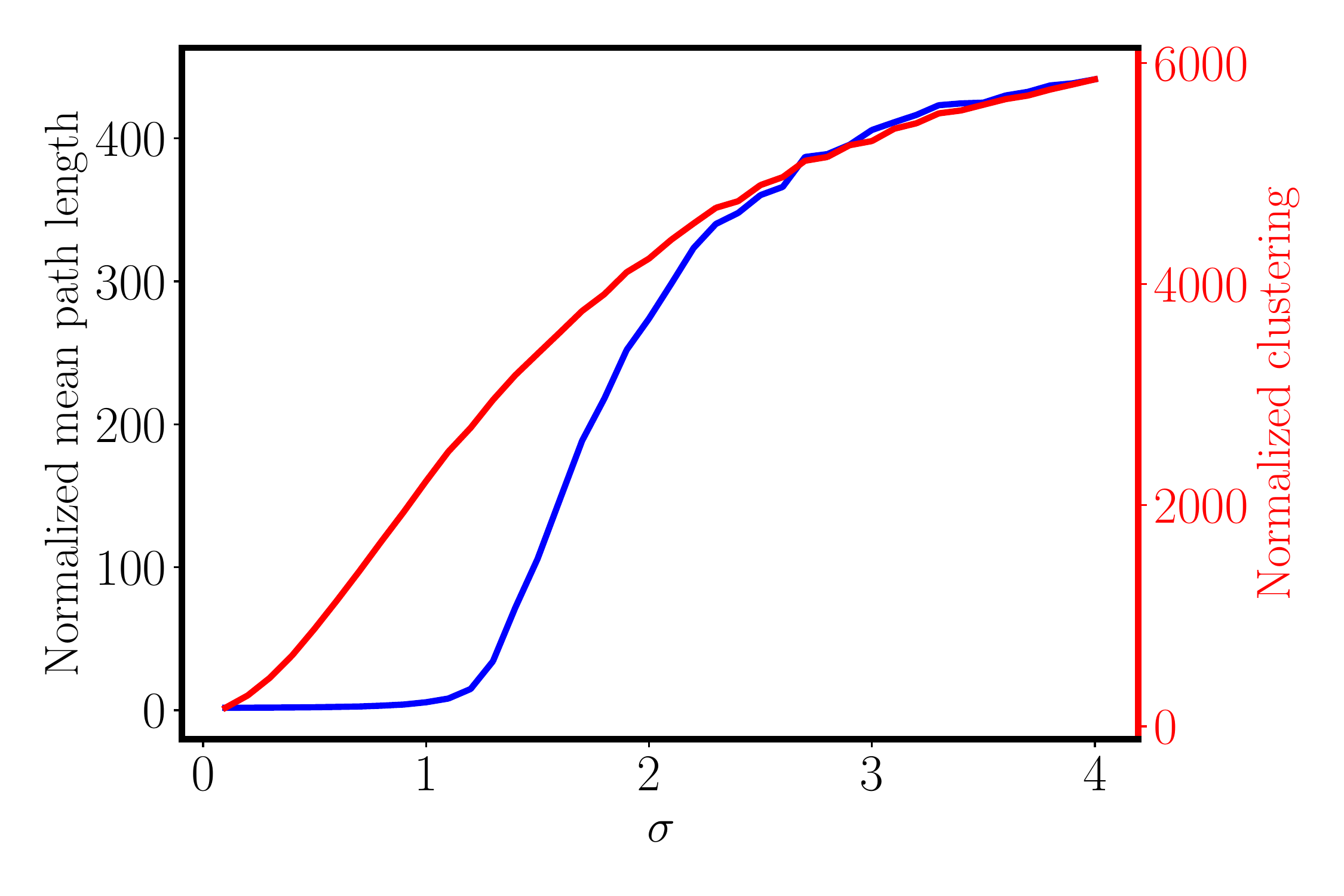}
\caption{Normalized mean path length $\ell(\sigma)/\ell_0(\sigma)$ (bottom line, in blue) and clustering coefficient $C(\sigma)/C_0(\sigma)$ (top line, in red) of the model. Null model: theoretical expectation for ER network with equal $N$ and $N_E$ as the corresponding LRRR network. \label{fig:null}
 }
\end{figure}

In Fig.\,\ref{fig:SI1}(b) we show the number of triangles presents in the network ($N_T$) as a function of $\sigma$, from which the clustering coefficient is calculated \cite{watts1998collective,newman2011networks}. As it can be seen, the number of triangles diverges as $\sigma \to 0$ and vanishes as $\sigma \to \infty$, where the network becomes a 1D circular chain. 
Moreover, in the $\sigma\to 0$ limit, the total number of possible triangles (given by $\kappa$), diverges faster than $N_T$, and therefore the clustering remains small, as shown in the main text. 
In Fig.\,\ref{fig:null} we show the clustering coefficient and mean path length of LRRR networks normalized over the corresponding values of the null model given by random Erd\H{o}s-R\'enyi (ER) networks with the same size ($N$) and number of edges ($N_E=N\kappa$). As can be seen, LRRR networks always have higher clustering and average distance than equivalent ER networks. As $\sigma\to 0$, LRRR become increasingly random and the difference decreases.

\bibliographystyle{apsrev_titles}
\bibliography{bibliography}

\end{document}